\documentclass{SCIS2018}

\begin{document}

\ArticleType{INVITED REVIEW}
\Year{2018}
\Month{April}
\Vol{61}
\No{1}
\DOI{}
\ArtNo{}
\ReceiveDate{}
\ReviseDate{}
\AcceptDate{}
\OnlineDate{}

\title{DNA Computing for Combinational Logic}

\author[1,2,3]{Chuan ZHANG}{chzhang@seu.edu.cn}
\author[1,2,3]{Lulu GE}{}
\author[1,2,3]{Yuchen ZHUANG}{}
\author[1,2,3]{Ziyuan SHEN}{}
\author[1,2,3]{\\Zhiwei ZHONG}{}
\author[2,3]{Zaichen ZHANG}{}
\author[2]{Xiaohu YOU}{}

\AuthorMark{ZHANG C}

\AuthorCitation{ZHANG C, GE L L, ZHUANG Y C, et al}

\contributions{Chuan Zhang and Lulu Ge have the same contribution to this work.}

\address[1]{Lab of Efficient Architectures for Digital-communication and Signal-processing (LEADS)}
\address[2]{National Mobile Communications Research Laboratory, Southeast University, Nanjing {\rm 211189}, China}
\address[3]{Quantum Information Center of Southeast University}

\abstract{With the progressive scale-down of semiconductor's feature size, people are looking forward to More Moore and More than Moore. In order to offer a possible alternative implementation process, people are trying to figure out a feasible transfer from silicon to molecular computing. Such transfer lies on bio-based modules programming with computer-like logic, aiming at realizing the Turing machine. To accomplish this, the DNA-based combinational logic is inevitably the first step we have taken care of. This timely overview paper introduces combinational logic synthesized in DNA computing from both analog and digital perspectives separately. State-of-the-art research progress is summarized for interested readers to quick understand DNA computing, initiate discussion on existing techniques and inspire innovation solutions. We hope this paper can pave the way for the future DNA computing synthesis.
}

\keywords{Synthetic biology, DNA computing, DNA strand displacement reactions, chemical reaction networks, combinational logic}

\maketitle

%
%
%
%
%
%


\section{Introduction}\label{sec:intro}
\begin{figure}[htbp]
  \centering
  \includegraphics[width=0.45\linewidth]{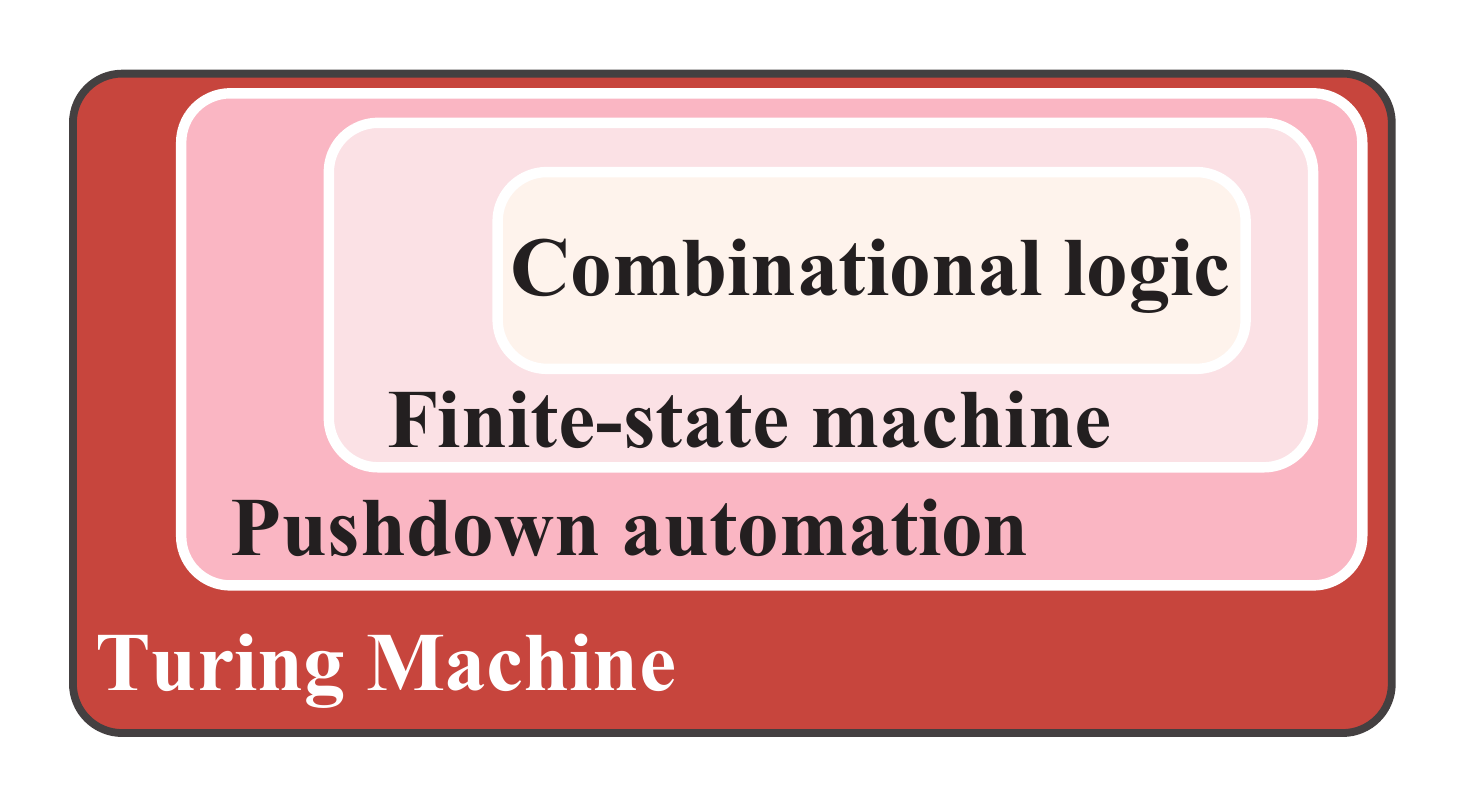}\\
  \caption{Automata theory \cite{comb}.}\label{fig:turing}
\end{figure}

\quad With the intense pursuit of More Moore and More than Moore \cite{mooreslaw2017,cnet2017,kish2002end,desai2016mos2}, other computing modes have emerged to meet people's ever-increasing demands on both computing speed and volume. Among all promising candidates, molecular computing, especially DNA computing, attracts extensive research interest for its massively parallel operations and high computational speed. With its final goal to operate as a Turing machine shown in Figure \ref{fig:turing}, DNA computing requires the synthesis of combinational logic to be the first step \cite{yahiro2016implementation,comb}. In the blue print, such as the field of synthetic biology \cite{khalil2010synthetic,siuti2013synthetic,andrianantoandro2006synthetic},  cells are employed as living devices to perform useful tasks like disease diagnosis and therapeutic drug release inside a body \cite{programmingcells2017}. However, the limited existing techniques leads to a current consensus---to programme bio-based modules with computer-like logic.




Due to its undoubted significance, research on combinational logic based on DNA computing has drawn a lot of attentions from both academia and industry. This timely special issue and this introductory overview paper summarizes main synthesizing techniques for combinational logic in DNA computing. We hope to help readers to learn and understand the cutting-edge progress from two perspectives---analog and digital. On this basis, more and more innovation solutions are hoped to be inspired to pave the way for DNA computing's realization.

The remainder of this paper is organized as follows. Section \ref{sec:dna} reviews DNA computing in the aspect of its history, property and application. Several preliminaries, such as the illustration of gene expression, are also given in this section for a good further explanation. Analog circuits design based on DNAs are described in Section \ref{sec:analog}. Section \ref{sec:digital} reviews the combinational logic from the perspective of digital circuits, strongly appealing a trend which combines analog and digital together. In closing, Section \ref{sec:future} concludes the entire paper and points out the possible future development of combinational logic in DNA computing.

\section{DNA computing}\label{sec:dna}
\quad As long ago as the 1950s, Feynman, who was the founder of nanotechnology, envisioned that we could manufacture an object maneuvering what we want on a very small scale \cite{feynman1960there}. However, it was not until 1994 that this notion of performing computations at a molecular level was realized by Adleman \cite{adleman1994molecular}. Due to his seminal work on computing with DNA, the field of DNA computing was initially developed. Since then, a tremendous amount of work has been done in this intriguing and nascent field, theoretically and experimentally.

\subsection{DNA vs silicon}
\quad The basic idea in DNA computing is from silicon to carbon \cite{paun2005dna}. With digital switching primitives replaced by DNA molecules, the computational substrate, which may be defined as ``a physical substance acted upon by the implementation of a computational architecture'', is changed from silicon to DNA \cite{amos1999theoretical}. For the traditional silicon-based computer with the \textit{von Neumann architecture} \cite{von1993first}, there exist two main barriers which prevent it going further on a small scale. \textcolor{blue}{\textit{1).}} The inherent \textit{von Neumann bottleneck} caused by the machine architecture \cite{backus1978can}. Literally, this \textit{von Neumann bottleneck} refers to a connecting tube that can transmit a single word between the central processing unit (CPU) and the memory. A data traffic of a problem is inevitable when pumping vast numbers of words back and forth in this bottleneck, which seriously limits the effective processing speed of CPU. As data size increases, the bottleneck has become more of a problem. It is also a limiting factor on the speed of the whole computer. \textcolor{blue}{\textit{2).}} The limitation derived from the nature of the computational substrate, silicon. At one time, $5$ nm silicon technology node was once anticipated by some experts to be the end of Moore's Law, because quantum tunnelling will occur in transistors smaller than $7$ nm scale \cite{mooreslaw2017,cnet2017,kish2002end}. Consequently, the unwanted interference severely affects the overall performance.

To overcome these two barriers of traditional silicon-based computers, some people are seeking for other computing modes. Their efforts are made mainly for two things. \textcolor{blue}{\textit{1).}} Break through our word-at-a-time thinking style given by the \textit{von Neumann bottleneck} and seek for non-von Neumann principles for new computer architectures. \textcolor{blue}{\textit{2).}} Prove the superiority of new implementations over the conventional one and then investments should be considered to offer an economic basis for further developments. Among all the promising candidates, include quantum, optical, and DNA-based computers, this paper concerns DNA computing for its high-density storage and vast parallelism.

\subsection{Pros. and cons.}
\subsubsection{Benefits of DNA computing}
\quad The following are main advantages of DNA computing:

\begin{itemize}
\item \textbf{Vast parallelism.} The most appealing advantage of DNA computing is its ability to address millions of operations in parallel \cite{deaton1997dna}. If forced to behave sequentially, DNA computer will lose its edge \cite{tagore2010dna}.
\item \textbf{Gigantic memory capacity.} One kilogram of DNA could meet the storage requirement from the whole world, under the premise of packaging the information as densely as it is in the genes of \textit{Escherichia coli}. With over $100$ years of data retention, the data density of bacterial DNA is estimated around $10^{19}$ bits per $cm^3$ \cite{digitalDNA}.
\item \textbf{Powerful computational capacity.} The seminal work done by Adleman uncovers the ability of DNA computing to solve NP-complete and other difficult computational problems, which are intractable for conventional computers \cite{hameed2011dna}.
\item \textbf{Easy availability.} DNA, as a non-toxic material, is easy to be fetched from the nature \cite{dnaSab}, as long as there are cellular organisms. This large supply of DNA makes it a cheap resource \cite{saxenaintroduction,kumar2015proper}.
\end{itemize}

\subsubsection{Limitations for DNA computing}
\quad Current limitations on DNA computing are listed below:

\begin{itemize}
\item \textbf{One-time ``instance" computers.} The current DNA computer is good at one instance of one specific problem, and typically performs computations only one time \cite{tagore2010dna}.
\item \textbf{Long time consumption.} Since massive and messy, reactions themselves are very slow, requiring time measured in hours or days for the final computational result \cite{oaodbc}.
\item \textbf{Limited synthesis technique.} DNA computing requires DNA synthesis and sequencing. If the high-quality synthesis is considered with no defect, the current practical length of synthetic oligonucleotides do not exceed $150\!\!-\!\!200$ bases \cite{ma2012dna,bornholt2016dna,hughes2017synthetic}. This hinders the capability of the constructed DNA-based computer to handle problems of larger size.
\item \textbf{Error prone characteristic.} DNA synthesis is prone to errors \cite{deaton1997dna}. Even the seemingly simplest mixing operation can sometimes cause problems.
\item \textbf{Resource-intensive problems.} Employing the computational paradigm developed by Adleman, even for some relatively simple problems, unrestricted solution space and impractical amounts of memory may be required \cite{amos1999theoretical}. 
\item \textbf{No visual output mechanism.} Since envisioned by \cite{paun2005dna}, the designed DNA-based computer is in the form of tube with liquid. The final result of this liquid computer is difficult to be observed by naked eyes. There should be a mechanism to deeply analyze and visually show the computational results.
\end{itemize}

\subsection{What does a DNA computer look like?}
\quad To the naked eye, the DNA computer looks like clear water solution in a test tube. No mechanical device there. A single drop of water surprisingly contains a trillion bio-molecular devices. Results are analyzed using a technique that allows scientists to see the length of the DNA output molecule \cite{tagore2010dna}, rather than show up on a computer screen.
%
%
%
%

\subsection{Application: what can DNA-based computers do?}
\quad Undoubtedly, DNA computer can do almost whatever the silicon-based one can. Consider the fact that silicon-based computer operates in a linear manner, while DNA computer is good at parallel processing. If DNA computer is forced to do as sequentially as silicon one, hours even years might be required to wait for a final result. This can not be tolerant. Therefore, to develop in their best way, the domain of DNA computers possibly covers the problems whose algorithms can be highly parallelized, while the speciality of silicon-based computers resolves those problems with sequential algorithms.

\textbf{Disease treatment.} DNA computer can help with disease treatment. Early in 2004, researchers constructed a simple molecular-scale DNA computer, coupled with both input and output modules \cite{benenson2004autonomous}. It is theoretically capable of correct cancerous diagnosis and then releasing an anti-cancer drug.

\textbf{DNA$^{2}$DNA applications.} This application is short for DNA \textit{to} DNA computations, where conventional computers cannot outshine DNA computers \cite{LandweberLR97}. Its key idea is to employ DNA computation with certain \textit{known} DNA strands to operate on \textit{unknown} pieces of DNA. After the unknown pieces of DNA are re-coded, reasonable computations can be performed on them to solve many interesting problems. Unlike ``classic'' DNA computations, these DNA$^{2}$DNA computations do not require the operations to work perfectly. Thus the results are much more error tolerant.

\textbf{DNA cryptography.} Due to the size of one-time-pad, practical applications of cryptographic systems based on one-time-pads are limited in traditional electronic media. Owing to the encryption of natural DNA and artificial DNA encoding binary data, DNA computing can be utilized in the cryptography field \cite{watada2008dna,gehani2003dna}.

\subsection{Several categories}
\begin{figure}[htbp]
  \centering
  \includegraphics[width=0.75\linewidth]{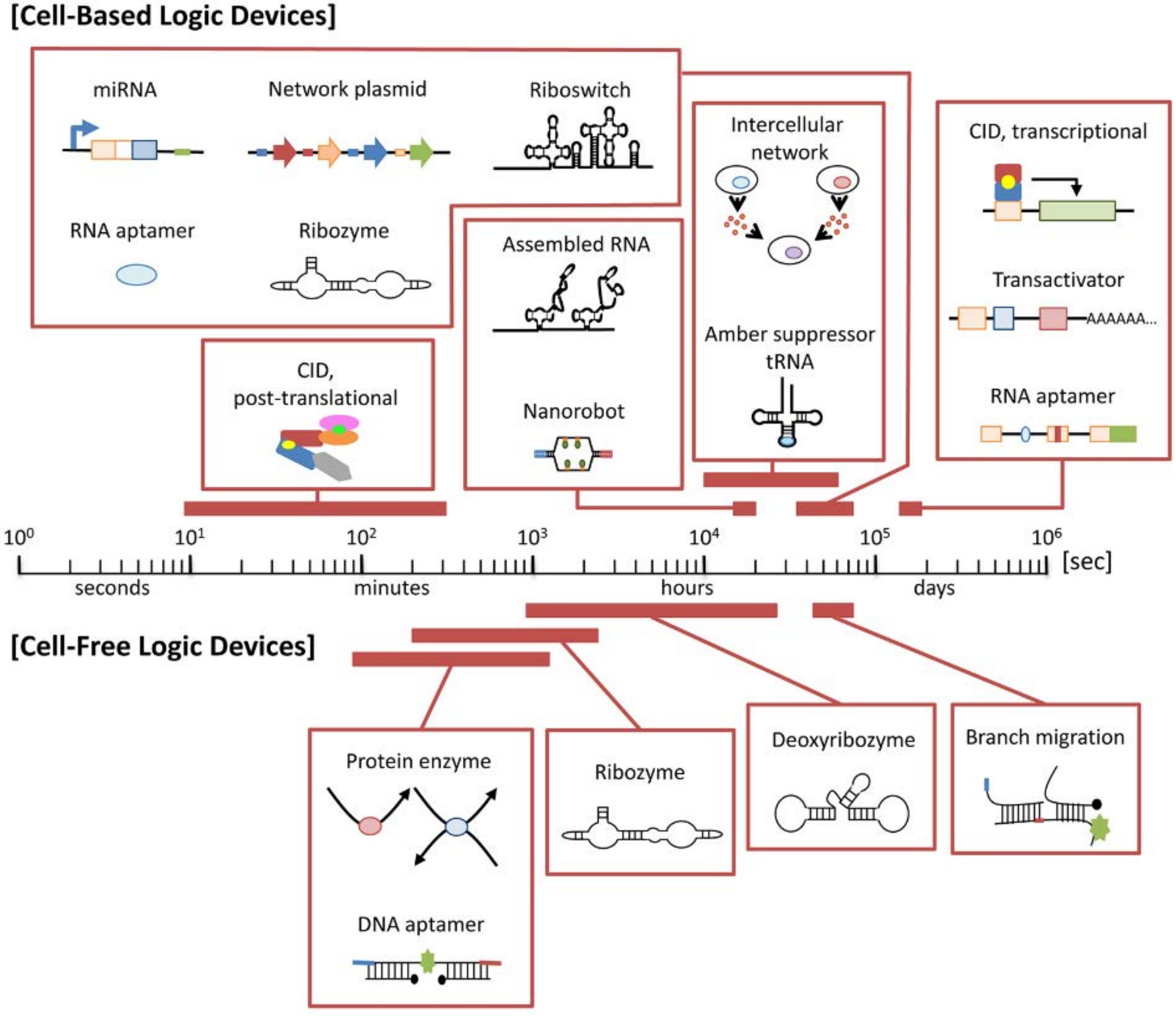}\\
  \caption{Representative time-scale for the activation time of biomolecule-based logic devices in \cite{miyamoto2012synthesizing}.}\label{fig:timescale}
\end{figure}

\quad In \cite{miyamoto2012synthesizing}, biomolecule-based Boolean logic gates fall into two categories: within cell and cell-free systems. According to the different substrates, they can also be categorized into two classes: Nucleic acid-based computation and protein-based counterpart. Based on whether there is enzyme or not, Nucleic acid-based computation includes both enzyme-based one and enzyme-free one. Figure \ref{fig:timescale} vividly and precisely displays the time-scale of activation time of cell-based and cell-free logic devices.

\subsection{Chemical reaction networks}
\quad Chemical reaction networks (CRNs) have been used as a formal language for modelling molecular systems, which comprise a set of reactants, products, and reactions. Moreover, CRNs could be employed to not only analyse the existing chemical systems but also construct new chemical systems in a convenient way, without considering the physical substrate. Using CRNs, researchers have designed novel molecular devices that achieve functions originally in electronic field, e.g., combinational logic, sequential logic, discrete-time signal processing computations, and arithmetic elements \cite{jiang2013digital,zhang2015karnaugh,ge2017formal,zhang2017clock,zhang2018ldpc,zhang2018mealy,zhang2018pwm,zhang2018analog,zhang2018stochastic,zhang2018neuron,zhang2018arithmetic,zhang2017sequential,zhang2017markov,zhang2017cnn,zhang2016stochastic,zhang2015clock,jiang2011synchronous,salehi2014asynchronous,senum2011rate}. CRNs are always modeled in terms of mass-action kinetics \cite{erdi1989mathematical,horn1972general}, which are demonstrated by ordinary differential equations (ODEs) \cite{howard2009analysis,sgb,strogatz2014nonlinear,zauderer2011partial,hale2013introduction}. Therefore the ODE-based simulations are usually used to validate the CRN-based design. After the CRN-based design has been validated through simulations, engineering materials are required to physically implement the molecule devices. So far, DNA has been considered as an ideal engineering material for CRN-based molecular devices mainly due to the following reasons: \textit{\textcolor{blue}{1).}} DNA is a critical part of \textit{the central dogma} \cite{crick1970central}, and as well the genetic material of most creatures. Thus, DNA devices could interface with a living body without any rejection reaction that might be caused by silicon-based devices. \textit{\textcolor{blue}{2).}} Diverse reaction mechanisms could be easily realized by simple DNA sequence design because DNA strictly conforms to the Watson-Crick complementarity principle. This shows the facility and flexibility of CRN implementations using DNA. \textit{\textcolor{blue}{3).}} Literature \cite{soloveichik2010dna} theoretically proves that any CRN could be mapped into DNA strand displacement reactions, as long as it only contain bimolecular and unimolecular reactions. This makes CRN-based designs promising for biomedical applications.

\subsection{DNA strand displacement reactions}

\quad Short for toehold-mediated branch migration and strand displacement reactions, DNA strand displacement reactions can be viewed as an approach to physically perform complex computations or behaviors. As a kind of enzyme-free system, these reactions exhibit lower catalytic speedup than those catalyzed by enzymes, resulting in a long time consuming process for the final results. Driven by the free energy, DNA strand displacement reactions are limited by the initial amount of reactants. This means that no more information processing can be done as long as the system reaches an equilibrium state\cite{zhang2011dynamic}. DNA strand displacement reactions can be cascaded, which indicates that the output of one reaction can be used as an input for the downstream reaction. This feature enables DNA strand displacement reactions to scale up and perform more complex behaviors. Exclusion of limited lifetime and long time consuming, the prediction of nucleic acid hybridization and strand displacement kinetics reveals a control over DNA strand displacement reactions design. This control is also attested by an observation that the rate of strand displacement reactions can be quantitatively controlled over a factor of $10^6$ by varying the strength (length and sequence composition) of toeholds \cite{zhang2011dynamic,zhang2009control}. Therefore, the ever-increasingly important sequence design motivates the development of its automated software, for example, the DNA Strand Displacement (DSD) \cite{phillips2009programming}. Empirically, to balance maximizing sequence space with minimizing crosstalk, it is useful to employ a three letter alphabet of $\{A, T, C\}$ to design the input and output strands, because $G$ is known to be the most promiscuous nucleotide for its strongest base pair formation and its strongest mismatches \cite{zhang2011dynamic,santalucia2004thermodynamics}. More importantly, a conspicuous feature that can be further exploited is that DNA strand displacement reactions can be programmed by chemical reaction networks (CRNs) \cite{shapiro2013dna,soloveichik2010dna}. This facilitates their design process of complex behaviors using CRNs as a programmable language. 

\begin{figure}[H]
  \centering
  \includegraphics[width=0.9\linewidth]{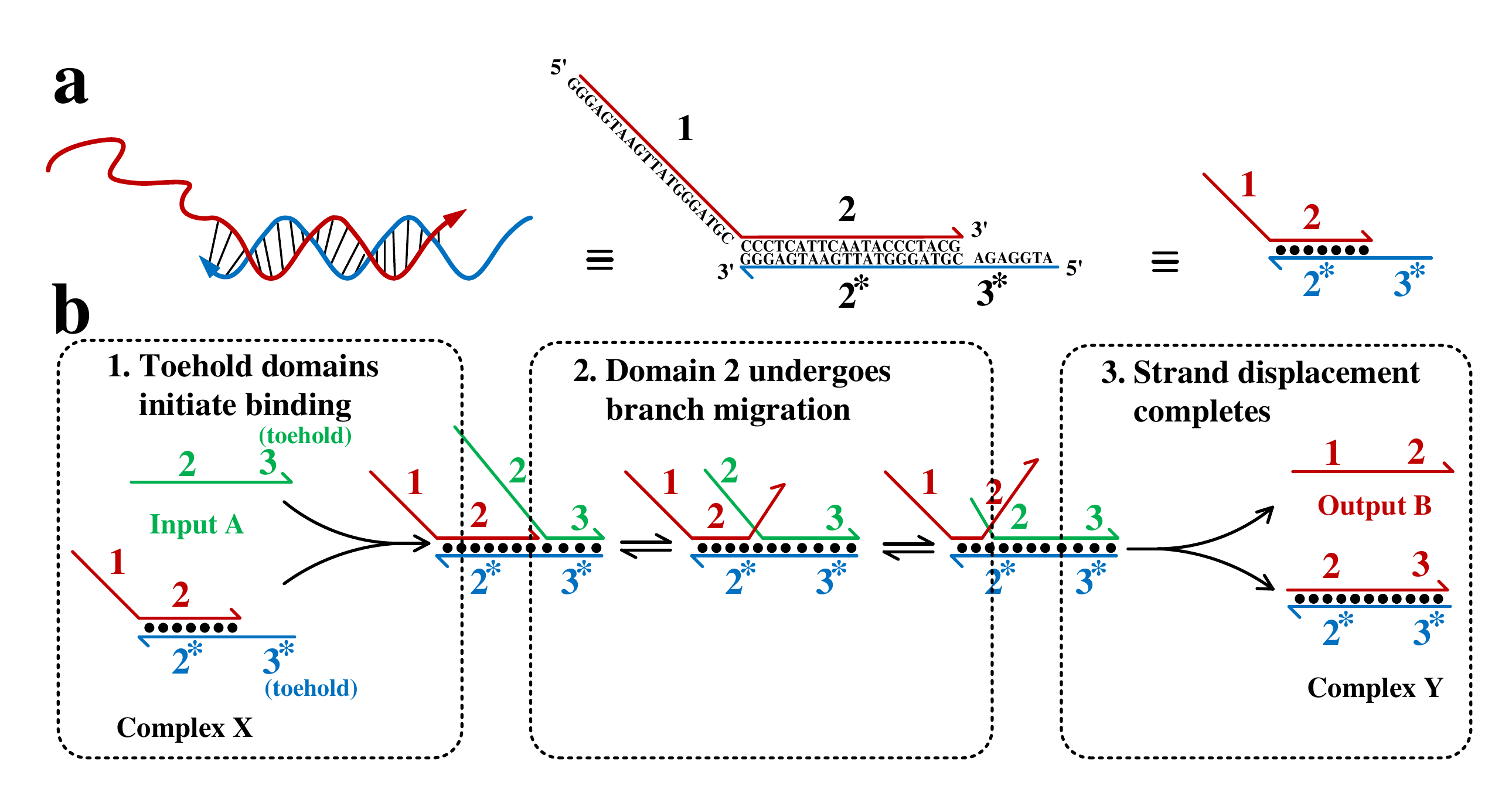}\\
  \caption{DNA strand displacement overview in \cite{zhang2011dynamic}.}\label{fig:radix}
\end{figure}

Figure \ref{fig:radix} offers a basic DNA strand displacement overview. Panel (\textbf{a}) vividly shows the structure of a DNA double helix. Directional lines, with the hook denoting the $3'$ end, are typically used to represent DNA molecules. The numbers $1$, $2$, $3$ actually represent different domains with specific DNA sequences. Panel (\textbf{b}) gives an example of DNA strand displacement reaction, namely $X+A \to Y + B$. Note that both the input $A$ and the output $B$ are single-stranded DNA molecules, while complex $X$ and $Y$ are double-stranded ones. Once the toehold domains $3$ and $3^*$ bind together, the strand displacement reaction is triggered. After the domain $2$ undergoes branch migration, the strand displacement reaction is finished. For more details, please refer to \cite{zhang2011dynamic,zhang2009control,zhang2010dynamic}.

\subsection{Gene expression}\label{gene expression}
\begin{figure}[ht]
\centering
\includegraphics[width=0.68\textwidth]{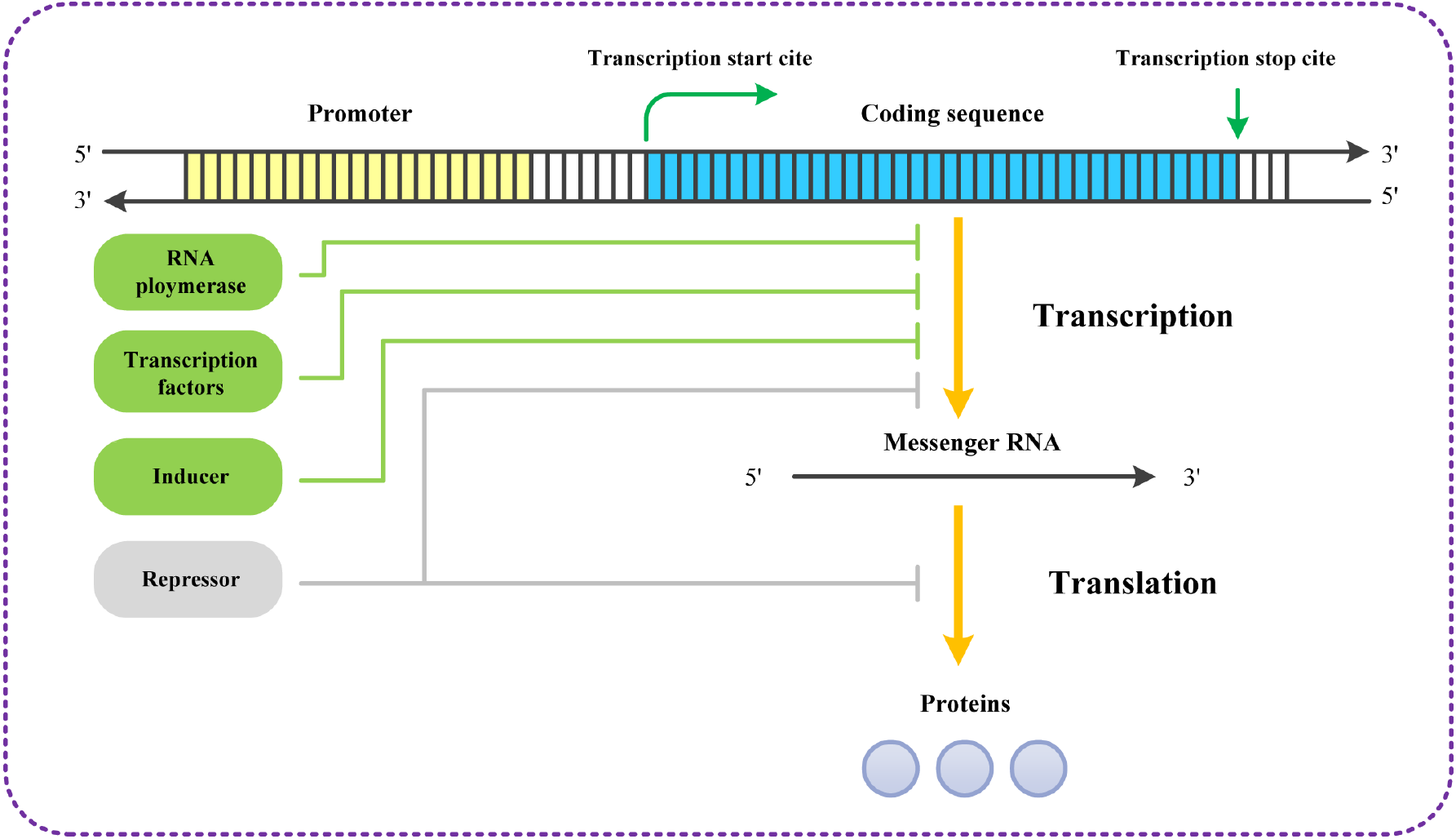}%
\caption{Basic molecular genetic processes.}\label{fig:gene}
\end{figure}

\quad Gene expression \cite{crick1970central, leavitt2004deciphering} is the process of transforming genetic information hidden in DNA sequences into functional proteins. It contains two steps: transcription and translation. Figure \ref{fig:gene} systematically illustrates a basic molecular gene expression. Both eukaryotes and prokaryotes are covered and merged. More specifically, transcription occurs first. In this step, the specific DNA sequence is copied into messenger RNA by the RNA polymerase (The green line indicates facilitation). Notably, the initiation of transcription is typically facilitated by a promoter (a special region of the same DNA strand) and transcription factors (proteins). After that, translation is triggered, in which ribosomes create polypeptides based on the messenger RNA. Then the polypeptides fold into functional proteins. In gene expression, the final output signals are the generated proteins; the input signals are molecules that can affect the gene expression. For example, a repressor could inhibit transcription or translation (The gray line indicates inhibition); the RNA polymerase and transcription factors initiate the gene expression as mentioned above; and an inducer could facilitate transcription by inhibiting the repressor or activating transcription factors. Different gene expressions could be viewed as different modules.

Actually, we categorise gene circuits into two types: analog and digital. In analog gene circuits, the concentrations of molecules above are used to indicate the values of input and output signals. Attention should be mainly paid to the functional relationships between the inputs and outputs. As for the digital circuits, logic values ``0'' and ``1'' are represented by the presence and absence of the molecules. Efforts should be focused on the establishment of the Boolean logic relationships between the inputs and outputs. The two types of gene circuits have one thing in common that they all have the potential to be cascaded due to the fact that the generated proteins of one gene expression might be used to adjust the expression of the other genes, functioning as its transcription factors, repressor or other relevant molecules.

\section{Analog logic}\label{sec:analog}
\quad One distinctive difference between analog DNA computation and digital DNA computation is that the input and output signals of analog DNA circuits are usually represented by the values of molecular concentrations, i.e., an analog DNA device is able to sense the concentrations of specific molecules in its environment and then produce certain concentrations of outputs through proper analog computations \cite{benenson2004autonomous}. Although digital circuits have taken a dominant place and achieved success in the field of electronics, analog circuits are more suitable and powerful than digital circuits in the field of DNA computation for the following reasons. \textcolor{blue}{\textit{1).}} Up to a given accuracy, analog circuits need much fewer gates to perform numerical computation than digital circuits \cite{sarpeshkar1998analog}. This resource-saving property of analog circuit is useful in living cells where resources are limited. \textcolor{blue}{\textit{2).}} Analog circuits are also more efficient than digital circuits in that cellular activities are analog computations in nature. Signals in cells are analog instead of digital \cite{sauro2013s}; the representation ``1'' and ``0'' in digital computation is too simple to indicate the effect of signals in cells.

To construct analog DNA circuits, one intuitive method is to translate traditional analog circuits into DNA circuits. Similar to the transformation from conventional digital electronics into digital DNA circuits, in which elementary DNA logic gates are synthesized to implement complex circuits, the transformation from analog electronics into analog DNA circuits should have focused on implementing DNA transistors. However, due to the fact that behaviour and property of a transistor are much more complicated and difficult to predict than those of a logic gate, the construction of a transistor is hardly analogous with behaviours of DNA strands, which only perform binding and unbinding operations in toehold-mediated strand displacement reactions. Therefore so far it is impractical to systematically synthesize DNA transistors to construct arbitrary analog circuits. One adopted alternative is to map analog electronics to analog DNA circuits at a behavioural level, i.e., to design DNA circuits achieving functions that originally implemented by silicon-based analog circuits. The following two subsections discusses: $(1)$ analog circuits only involving DNA strands, $(2)$ analog circuits involving gene expression.
\begin{figure}[ht]
\centering
\includegraphics[width=0.95\textwidth]{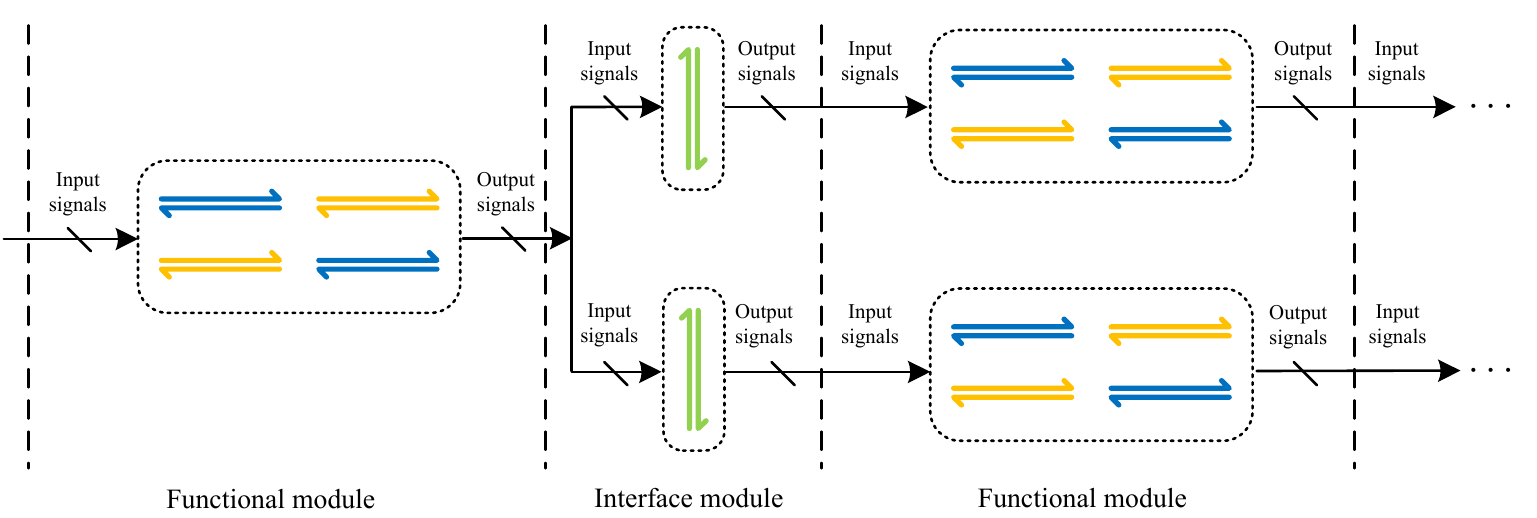}
\caption{Combination of different analog DNA circuits.}\label{fig:analog}
\end{figure}

\quad Typically, in analog DNA circuits, the concentrations of single-stranded DNA signals are utilized to indicate the values of input signals and output signals; double-stranded DNA complexes are used to process input signals and generate output signals. The analog DNA circuits are driven by toehold-mediated strand displacement reactions. Diverse functions have been achieved. For example, \cite{song2016analog} provides DNA devices to perform analog computation, including addition, subtraction, and multiplication based on DNA reactions. Using these elementary arithmetic modules, it also describes how to compute polynomial functions and even functions beyond the scope of polynomials. \cite{yordanov2014computational} utilizes DNA reactions to construct the feedback control circuit, which is common in analog electronics. \cite{chen2013programmable} employs three DNA reactions to synthesize a decision-making controller, which works on two types of DNA strands and is able to convert the one in lower initial concentration into the one in higher initial concentration. Theoretically, these analog DNA circuits could be connected to achieve more complex functions, since they all rely on the same mechanism: toehold-mediated strand displacement reactions. However, consider that the input and output signals in different circuits might be in different structures (the position of the toehold domain and the recognition domain), extra interface modules, composed of DNA reactions, are required to transform the output signals of one circuit into the input signals of another circuit without destructing the distinctness of signals as shown in Figure \ref{fig:analog}. Consider that toehold-mediated strand displacement reactions do not require any enzyme as catalyst, such circuits are assumed to be well applied in an enzyme-free environment, i.e., in \textit{vitro}.

\begin{figure}[ht]
\centering
\includegraphics[width=0.8\textwidth]{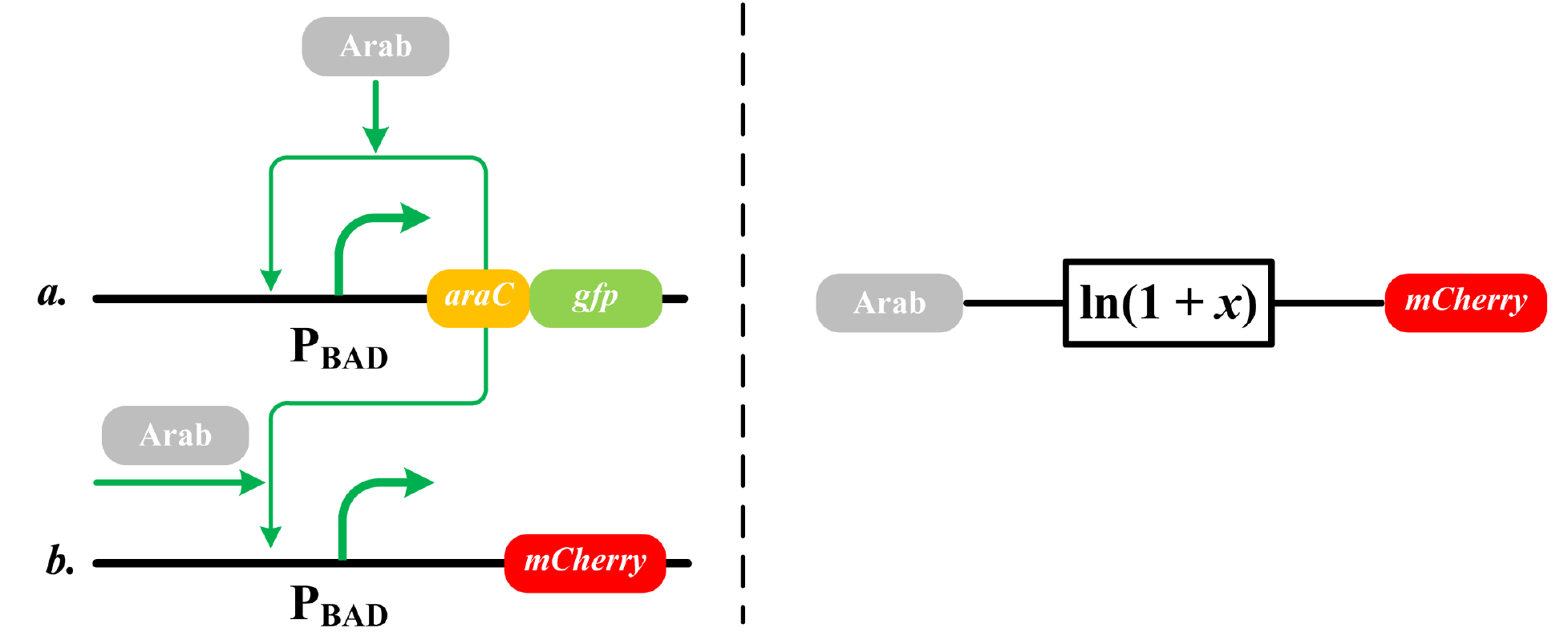}
\caption{Implementation of function $y=\ln(1+x)$ with gene circuits. Cited from \cite{daniel2013synthetic}.}\label{fig:ln}
\end{figure}

\quad Here we introduce analog gene circuits. Such circuits differ from analog DNA circuits in that the input and output signals of gene circuits are not DNA strands. Instead, in gene expression (DNA $\rightarrow$ RNA $\rightarrow$ protein), RNA polymerase, transcription factors, inducer, and repressor could be viewed as input signals; the final product-protein could be viewed as output signals. Gene circuits could be easily and directly synthesised in living cells. For example, an analog adder could be implemented by two inputs regulating the expression of a common output protein from independent genetic promoters, in which the two molecular fluxes will automatically add to generate the common output protein as the sum \cite{sarpeshkar2014analog}. In some cases, the generated proteins could also be used as transcription factors or repressors, thus making the gene circuits cascadable. For example, by using three gene parts, \cite{daniel2013synthetic} constructs the log-domain analog circuit in living cells, implementing function $y=\ln(1+x)$ as shown in Figure \ref{fig:ln}, in which $x$ refers to the concentration of the inducer (Arab) and $y$ refers to the concentration of the protein (\(mCherry\)). Specifically, in the first gene expression labelled $``a"$, the inducer (Arab) helps the transcription factor (\(araC\)) bind to the promoter (\(\rm P_{BAD}\)), so as to generate the output protein-\(araC\), which is also the transcription factor. Apparently, the generated proteins will further facilitate the gene expression. This could be viewed as a positive-feedback loop. In the second gene expression labelled $``b"$, the output proteins (\(araC\)) mentioned above are shunted away and also function as the transcription factors, generating the output protein (\(mCherry\)). Experiments in \cite{daniel2013synthetic} illustrate that the concentrations of Arab and \(mCherry\) conform to the function $y=\ln(1+x)$. In addition, \cite{sarpeshkar1998analog} has proven that in the field of electronics, the computation that is most efficient in its use of resources is the mixture of analog computation and digital computation. Thus we might assume that the same conclusion applies to DNA computation as we translate electronics into DNA circuits. Then the molecular analog-to-digital converters and digital-to-analog converters proposed in \cite{salehi2015molecular} will be critical in the mixture of the two forms of computation in the long run.


\section{Digital logic}\label{sec:digital}
\quad In the design of digital circuits, the Boolean values should be defined in an appropriate way and the bits of the binary system in the calculation should also be taken into account. The following existing literature show different angles of treating this issue.

An experimental study \cite{frezza2007modular} utilizes a similar method to represent Boolean values in logic circuits. This research is also based on the DNA strand displacement reactions. However, it is mainly about the experiments the researchers have done.  Digital logic gates, like AND, OR, NAND and XOR gates have been all accomplished in the experiments. To get larger scale logic circuits, the inputs and the outputs of different levels ought to be isolated until they have been displaced from the solid support by the appropriate combination of inputs. In the results, they monitor the relative concentration of fluorescence in certain DNA strands and define a threshold to classify the Boolean values. If the value is above the threshold, it is seen as ``1''. If not, the value ``0'' is achieved. However, the drawbacks of the method applied in the experiment is also obvious. All designs including the inputs and the outputs are analog approximations of digital circuits. Thus, imperfect strand displacement yields a minor loss in the input/output signal at each level of circuit operation, which fundamentally limits the size of circuits that can be constructed.

\begin{figure}[H]
\begin{minipage}[t]{0.51\linewidth}
\centering
\includegraphics[width =.5\linewidth]{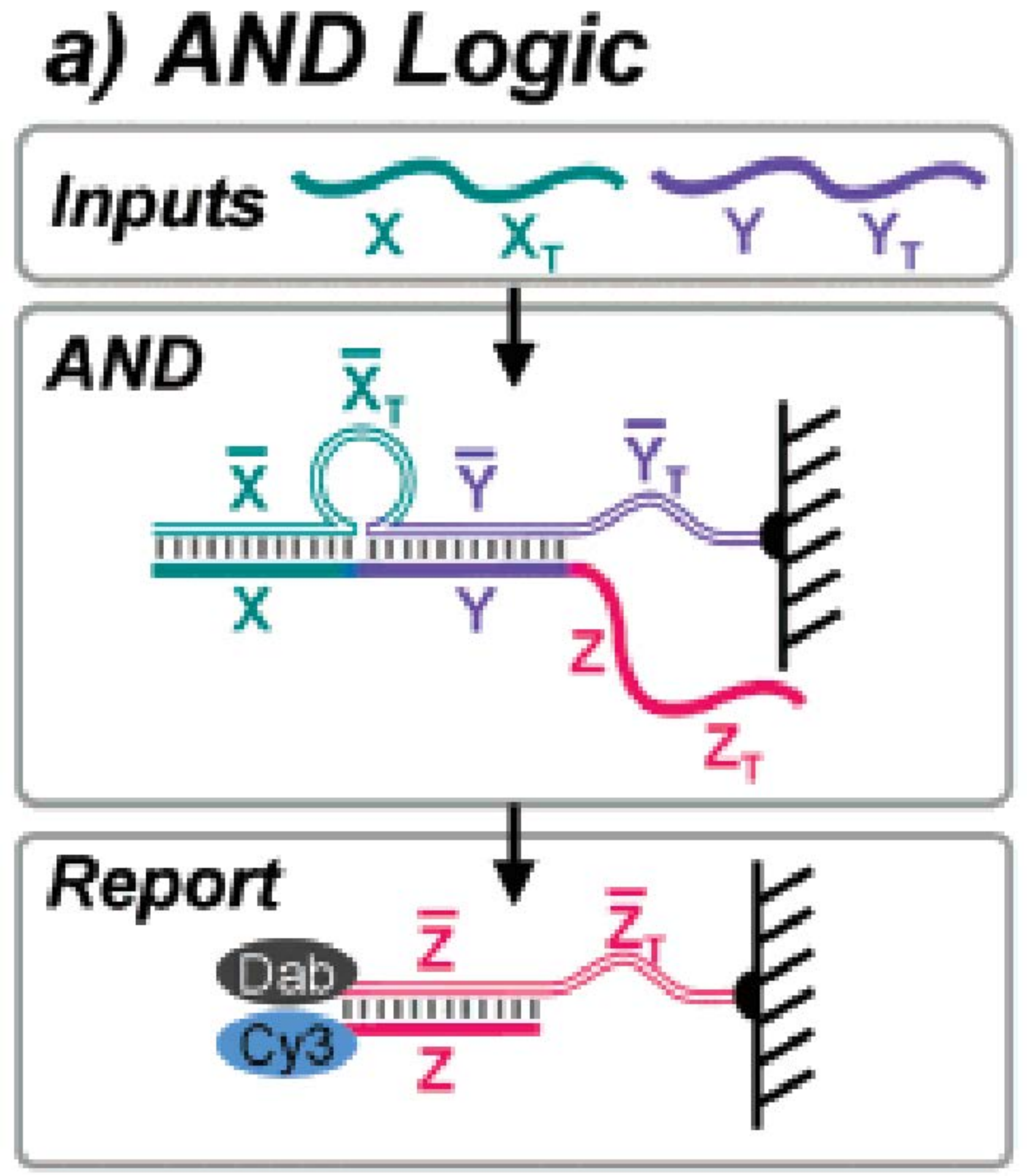}
\caption{Experiment method of AND logic in \cite{frezza2007modular}.}
\end{minipage}%
\begin{minipage}[t]{0.49\linewidth}
\centering
\includegraphics[width=.5\linewidth]{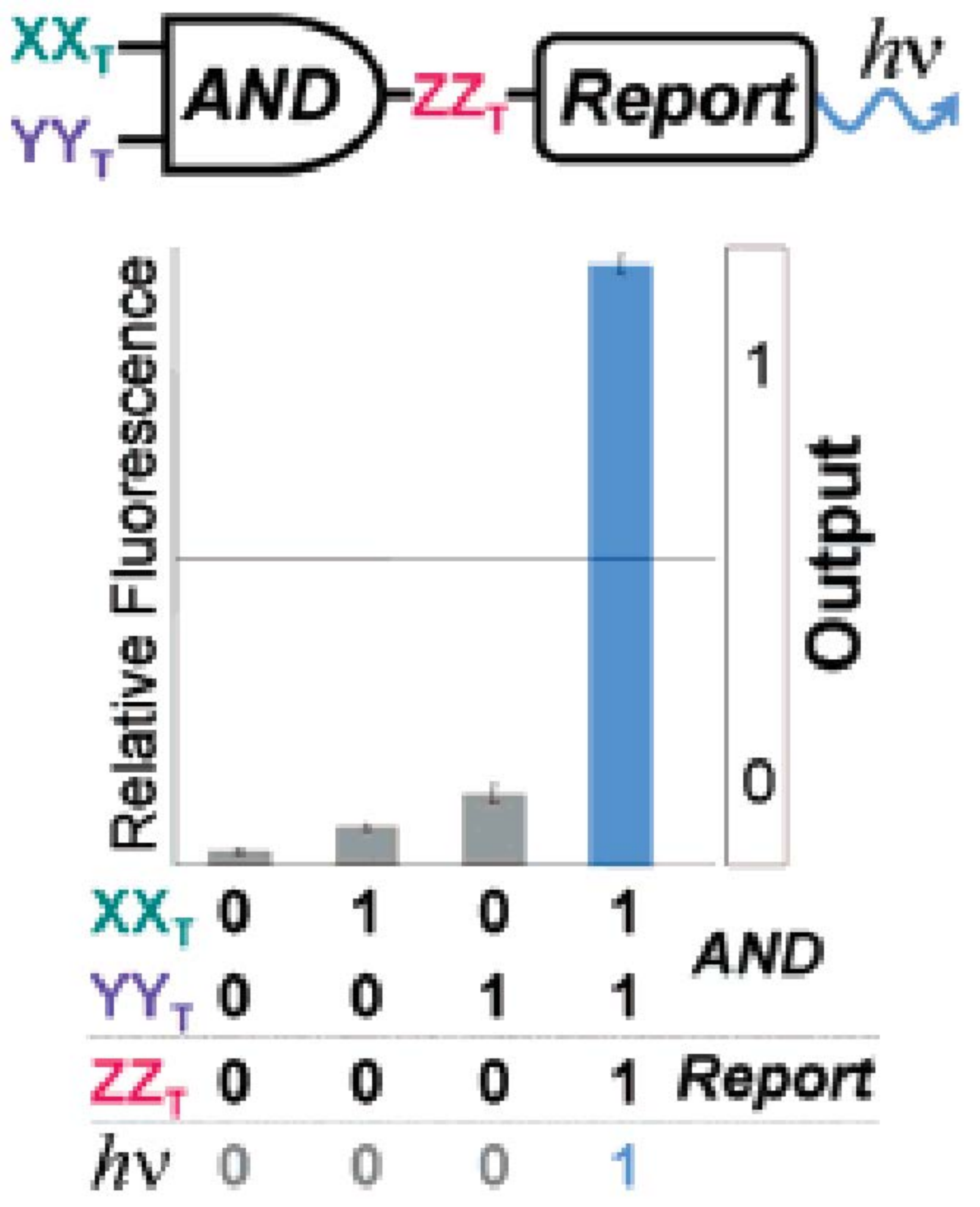}
\caption{Experiment result of AND logic in \cite{frezza2007modular}.}
\end{minipage}
\end{figure}

Single-rail representations like above are studied, while the dual-rails values can also make sense. The binary system is defined differently in \cite{chiniforooshan2010scalable}. Dual-rails value is utilized to represent the value ``0'' or ``1''. Given a certain reactant $w$, the existence of $0_w$ represents ``0'' while the existence of $1_w$ represents ``1''. All the designed chemical reactions obey this principle. The research employs this method to design some other parts of the digital circuits, for example the NAND gates, the fan-out gates and the signal restoration gates in chemical reactions. Also, the combination of these parts is studied. For experimental realization, an introduction of the reaction environment and the biology principles behind the reactions are listed as well. In the final parts of the article, the drawback of this kind of method is analysed that dual-rails make the detection of absence more complex. Aside from that, this method should make every effort to deal with some unideal inputs, like $0.99$ and $0.01$, which is often encountered in many researches in this area. The design cannot be renewable and the threshold is some kind of a too ideal thing in such methods.

Jiang Hua's research \cite{jiang2013digital} mainly talks about realizing digital modules with chemical reactions. The whole design is based on DNA strand displacement reactions. The digital values of ``0'' and ``1'' are represented by two different reactants, namely ``$X_0$'' and ``$X_1$''. The presence of ``$X_0$'' represents 0 and the presence of ``$X_1$'' represents 1. Similar to the method in \cite{chiniforooshan2010scalable}, the two reactants cannot coexist with each other. In this method, the principles of each bit calculating in the logic gates can be easily studied out. According to the principles, the basic digital modules are transplanted to the corresponding chemical reactions. Apart from the basic digital logic gates, both the D latches and D flip-flops are also designed with the same idea. To ensure the validity, the corresponding ODEs towards each reaction network are also listed. With these basic modules prepared, more complicated modules can be combined with each other, like full adders, half adders and linear feedback shift registers. So far, the general method of designing more complicated digital logic circuits is achieved. 


Besides the traditional designs, a specialized seesaw gate can be fabricated to combine all the logic gates together \cite{qian2011scaling}. Employing dual-rail logic to represent binary values, this seesaw gate has four basic components, including input, threshold, fuel and gate:output. A pair of seesaw gates, once cascaded, essentially perform AND or OR operations. To perform which operation depends on the initial concentration of the devised threshold. Combing with dual-rail logic, NOT operation is introduced. This means that any AND-OR-NOT circuit can be synthesized with seesaw gates using dual-rail logic, and translated into DNA. With this, designers can easily acquire the target DNA circuits following five sequential steps: a given AND-OR-NOT circuit, dual rail logic, seesaw gates, DNA domain and DNA sequence. Moreover, owing to the defined integrating gates and amplifying gates, the built logic circuits can support both multiple inputs and multiple outputs. Though use-once, the systematically synthesized digital circuits can be large-scale.

\begin{figure}[H]
\begin{minipage}[t]{1\linewidth}
\centering
\includegraphics[width =.5 \linewidth]{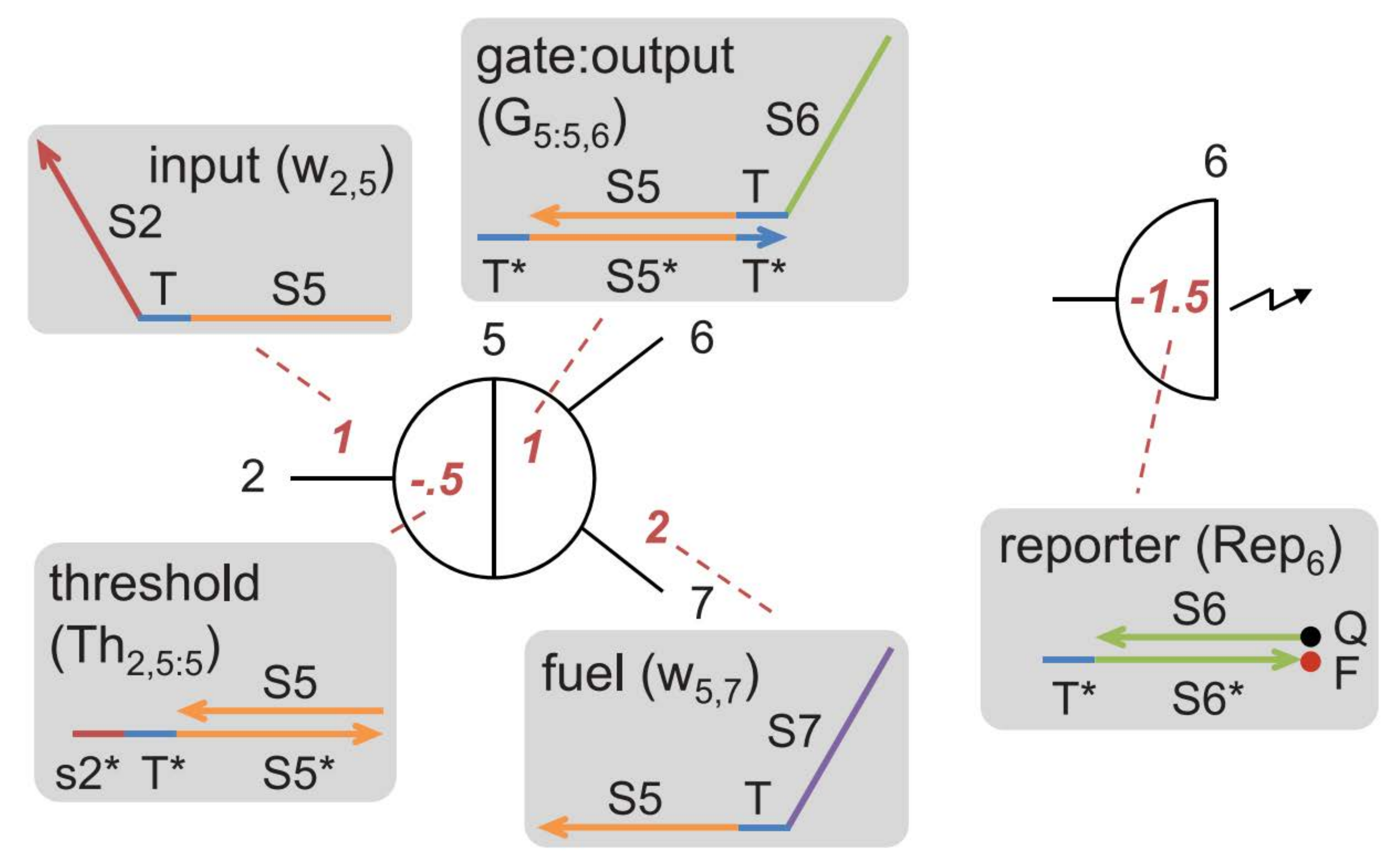}
\caption{A seesaw gate in \cite{qian2011scaling}.}
\end{minipage}%

\begin{minipage}[t]{1\linewidth}
\centering
\includegraphics[width=.77\linewidth]{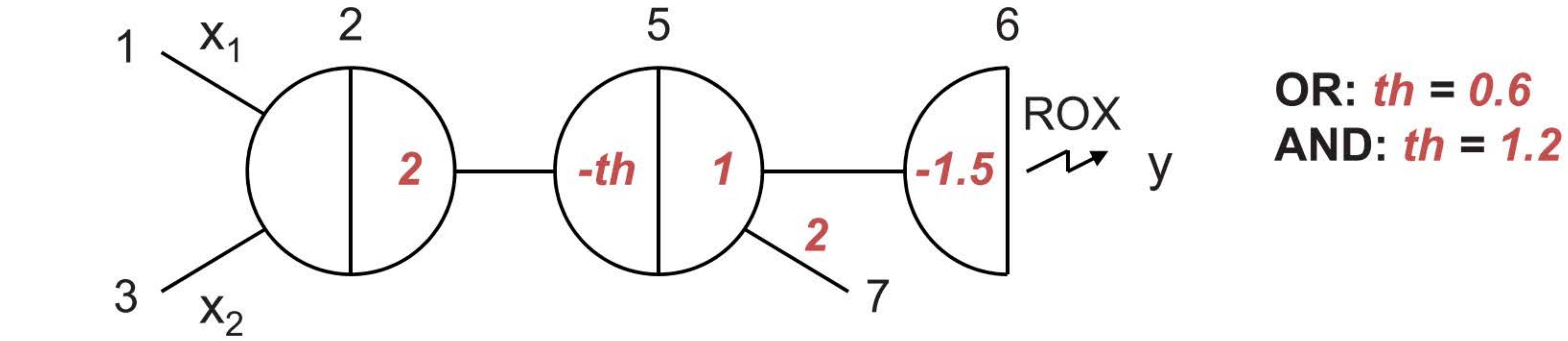}
\caption{Two cascaded seesaw gates with a reporter in \cite{qian2011scaling}.}
\end{minipage}
\end{figure}

\cite{nielsen2016genetic} designs a new hardware description language  for programming living cells. The researchers utilize the hardware description language Verilog to enable a user to describe a circuit function. Based on that, the new platform Cello \cite{Cello} uses the information to automatically design a DNA sequence encoding the desired circuit. The mechanism can be described as follows. To realize the final target, the Verilog text should be parsed first. According to the results of the parse, the circuit diagram can be created. For the reason that Cello designs circuits by drawing upon a library of Boolean logic gates, the next step is to assign those gates to accomplish the circuit design. Then we can achieve the conversion from traditional logic gates to genetic circuit DNA sequence. In the figure below, raised arrows are promoters, circles on stems are ribozyme insulators, hemispheres are ribosome binding sites (RBSs), large arrows are protein-coding sequences, and ``T''s are terminators. Part colors correspond to physical gates. Combining the gates together and making great efforts to isolate the different gates, the results of the insulated circuits are quite correct. Similar to some other researches, fluorescent protein is applied to present the final results. If the fluorescence is above the threshold, we can define that result as Boolean value ``1'', and otherwise as ``0''.

\begin{figure}[H]
\begin{minipage}[t]{0.45\linewidth}
\centering
\includegraphics[width =.9\linewidth]{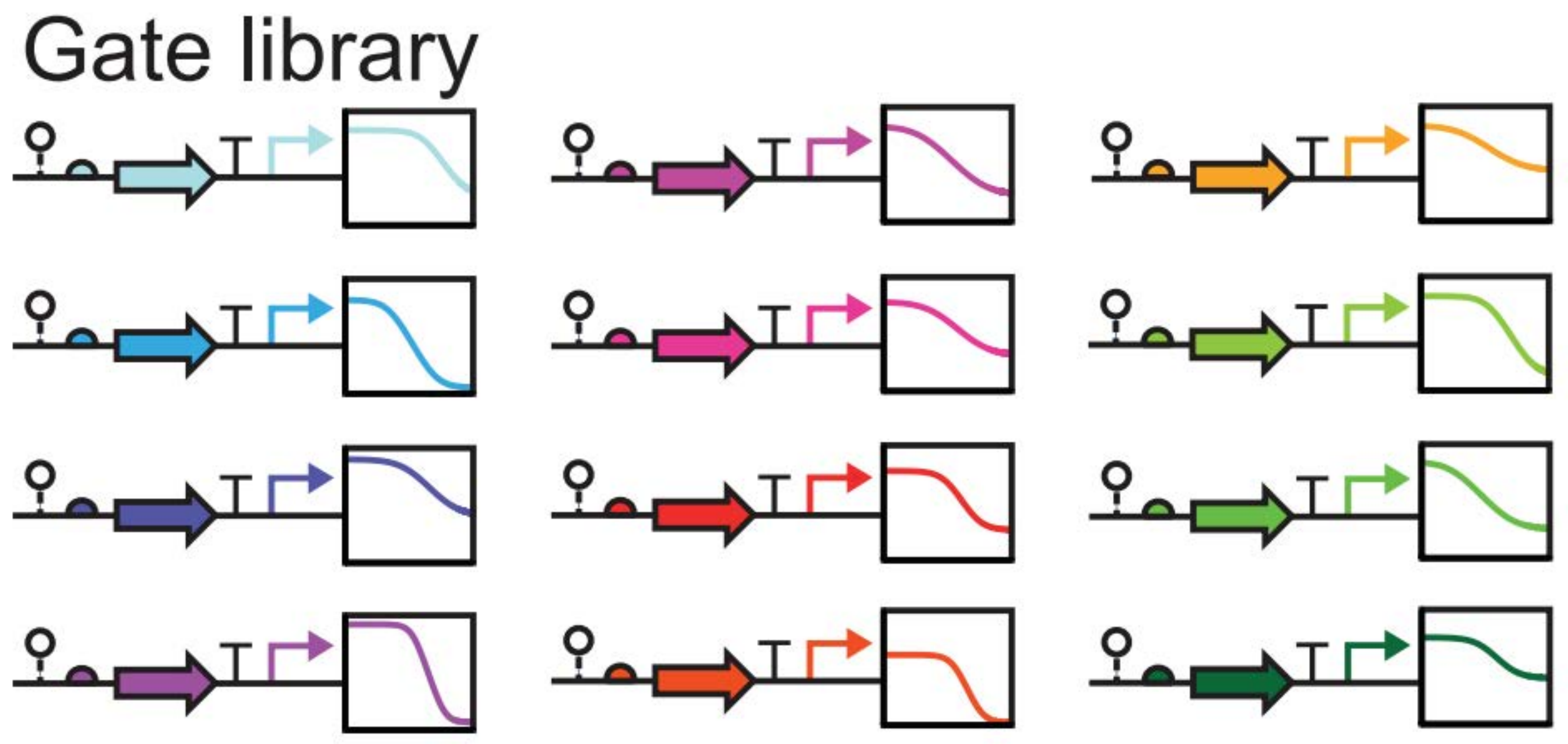}
\caption{The gate library in \cite{nielsen2016genetic}.}
\end{minipage}%
\begin{minipage}[t]{0.55\linewidth}
\centering
\includegraphics[width=.9\linewidth]{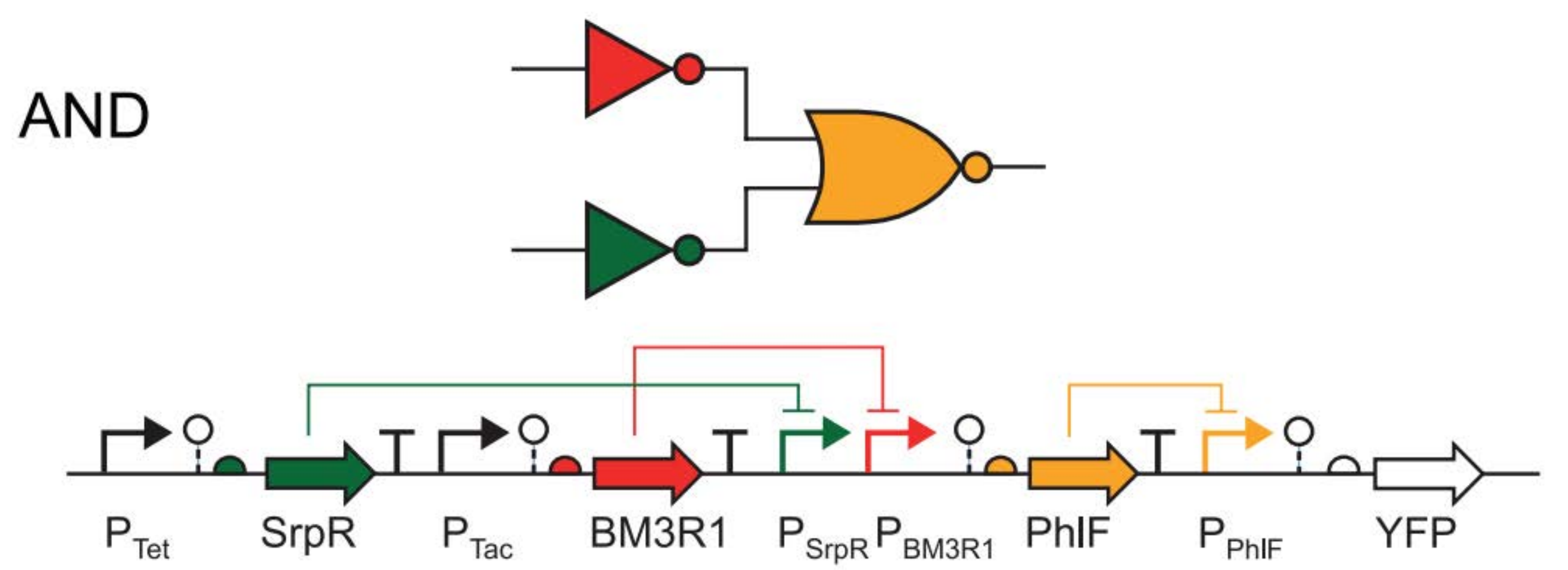}
\caption{Implementation of AND gate generated by Cello in \cite{nielsen2016genetic}.}
\end{minipage}
\end{figure}

Different strategies for creating synthetic digital and analog circuits in living cells are also reviewed in \cite{roquet2014digital}. In this review, the researchers mainly show their interest in digital memory, digital computation and multi-input digital logic. In the mentioned biological circuits, signals can be represented by chemical concentrations and digital signals that take on discrete values. The DNA-based memory storage has great potential in stability and scalability because of its high-parallelism, nevertheless it does not have an effective mean to maintain the states. To realize the circuits that is able to detect inputs, the biological environment and the endogenous cells can be used. And similar to \cite{frezza2007modular}, this paper also prefers to use a proper threshold to transfer the original analog signals to digital ones. The multi-input logic gates integrate all input signals combinatorially, which refer to digital computations without memory such that the output only depends on the present inputs.

\begin{figure}[H]
\begin{minipage}[t]{1\linewidth}
\centering
\includegraphics[width =.65\linewidth]{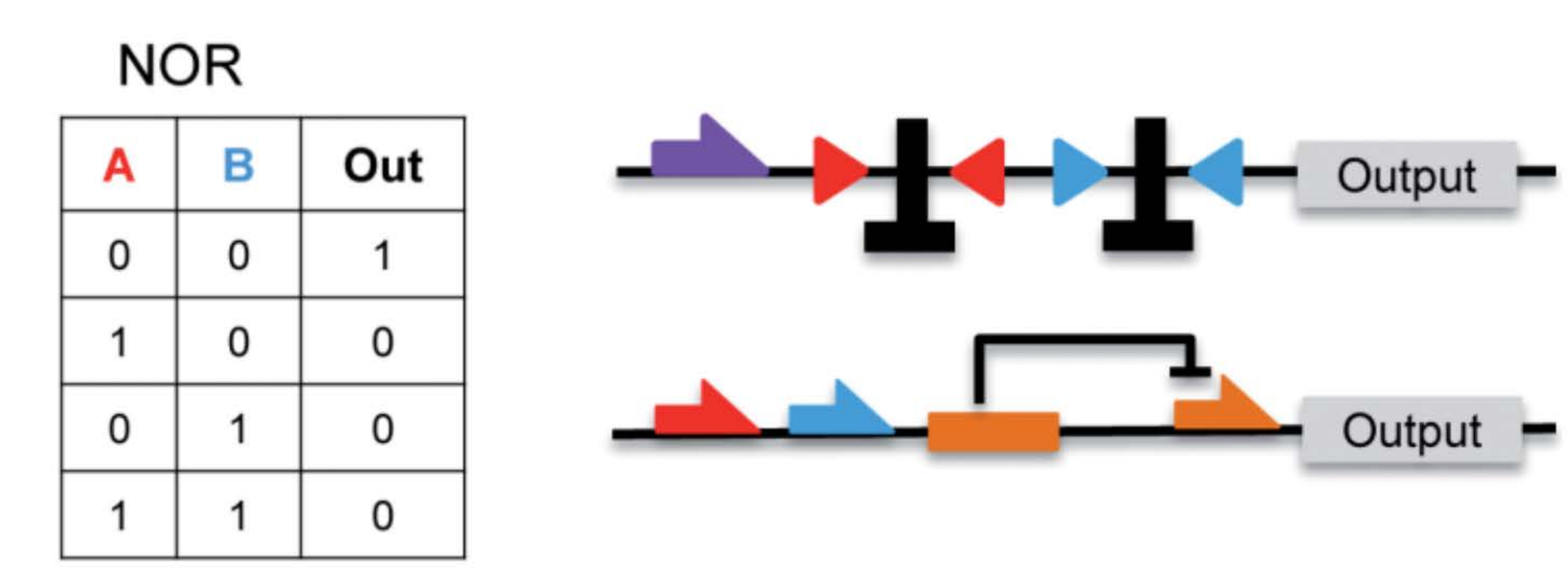}
\caption{An example of NOR digital logic gate synthesized in \cite{roquet2014digital}.}
\end{minipage}
\end{figure}

In the paper \cite{weiss2003genetic}, a more biological approach to establish genetic circuits is put forward. The aim of the article is to create synthetic gene networks that enable people to engineer cells with the same ease and capability with which people currently program computers and robots. And commonly, the scale is gradually increased. The research describes the genetic circuit building blocks, existing prototype circuits, genetic circuit design, cell-cell communication and signal processing circuits. The latter circuits are combined with the former smaller-scaled circuits. The main cellular blocks mentioned include components for intracellular computations (i.e., NOT and NAND) and devices for external communication (i.e., IMPLIES and AND). For a general study on this paper, the biochemical inverter can be taken as an example. Biochemical inversion uses the transcription and translation cellular processes. Ribosomal RNA translates the input mRNA into an animo acid chain, which then folds into a three-dimensional protein structure. When the protein binds an operator of the gene's promoter, it prevents transcription of the gene by RNA polymerase (RNAp). In the absence of the repressor protein, RNAp transcribes the gene into the output mRNA. The other parts like NAND gates apply the same method of transcription and translation cellular process. Then the larger-scale circuits can be established based on these basic genetic circuits.

\begin{figure}[H]
\begin{minipage}[t]{1\linewidth}
\centering
\includegraphics[width =.76\linewidth]{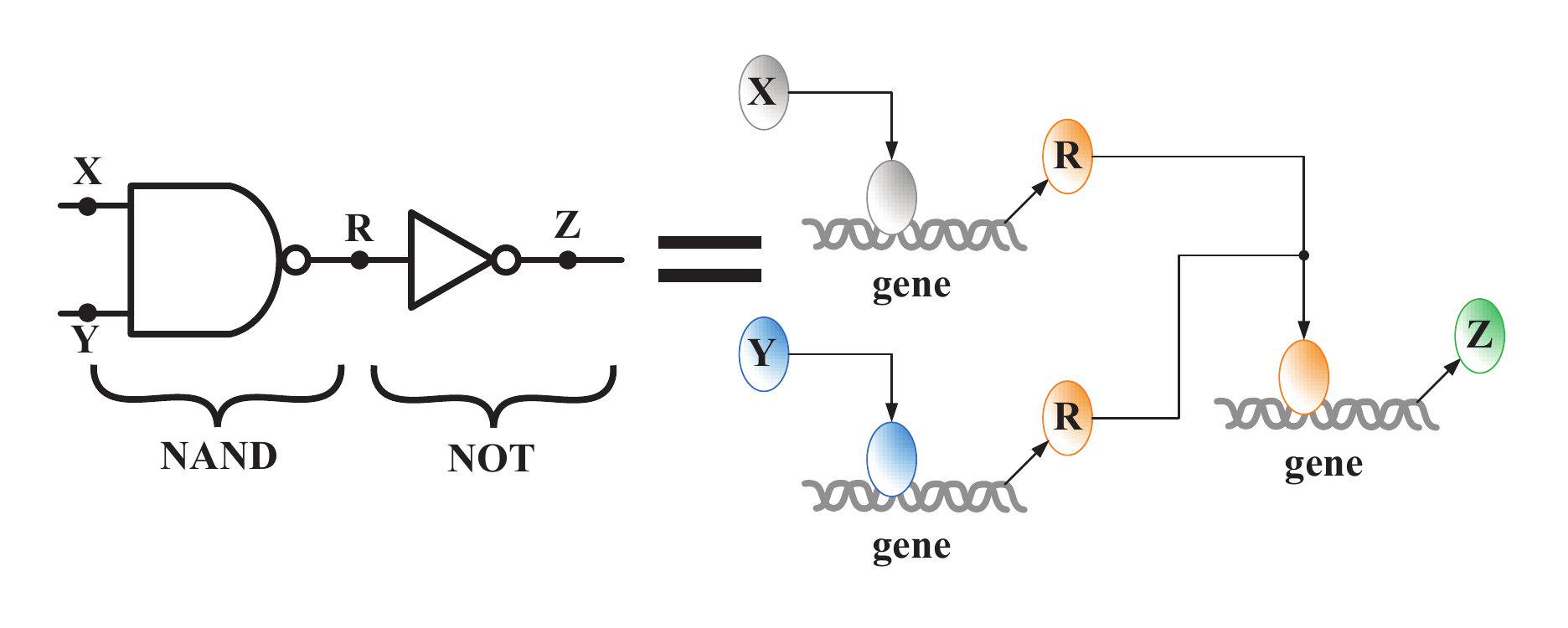}
\caption{An example of NOR digital logic gate synthesized in \cite{weiss2003genetic}.}
\end{minipage}
\end{figure}

\begin{figure}[H]
\begin{minipage}[t]{0.55\linewidth}
\centering
\includegraphics[width =.9\linewidth]{x3_nw2.pdf}
\caption{An approach for a $3$-input gate in \cite{ge2017formal}.}
\label{fig:ge1}
\end{minipage}%
\begin{minipage}[t]{0.45\linewidth}
\centering
\includegraphics[width=.7\linewidth]{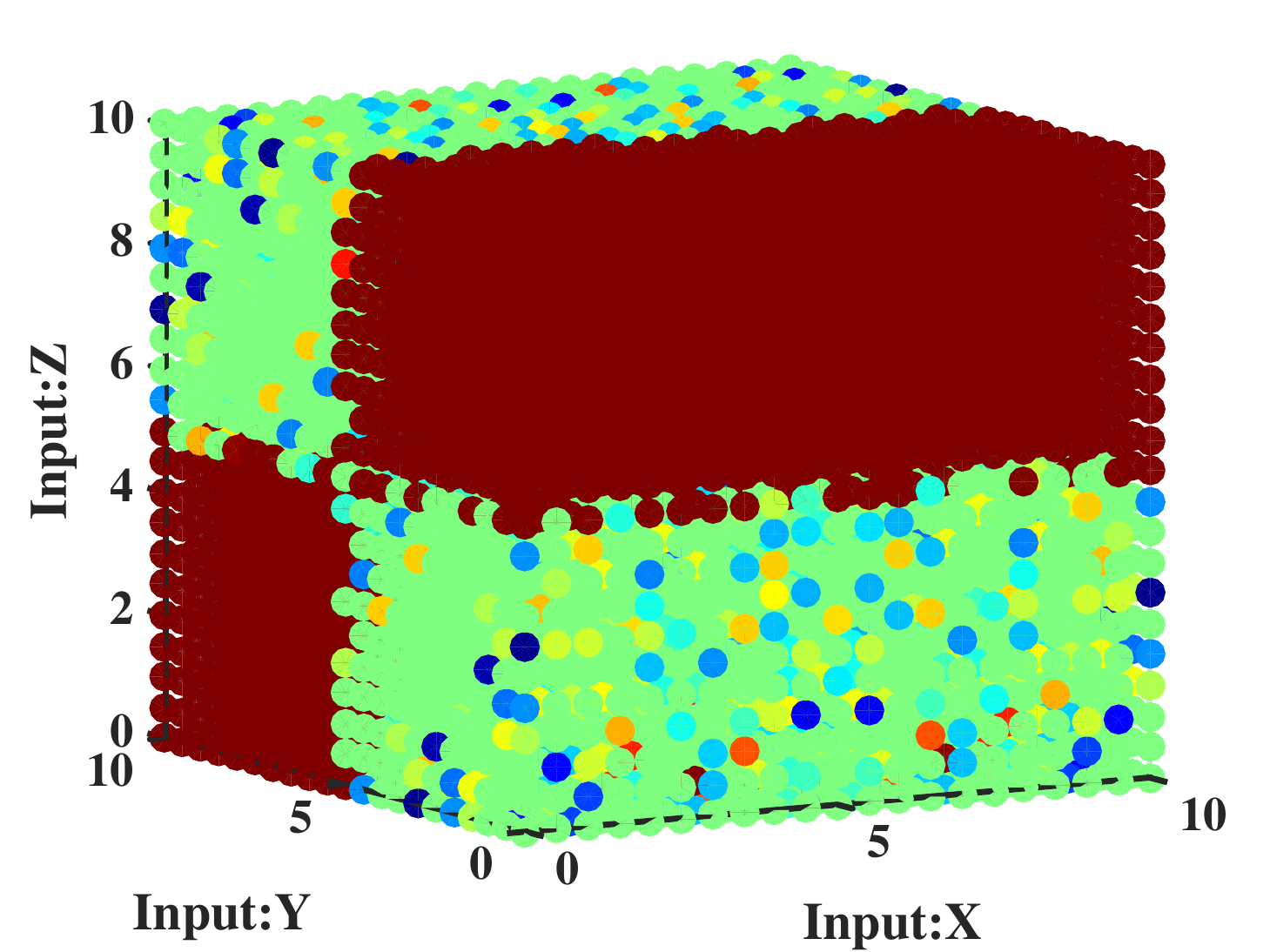}
\caption{Simulation of this $3$-input gate in \cite{ge2017formal}.}
\label{fig:ge2}
\end{minipage}
\end{figure}

In paper \cite{zhang2015karnaugh}, it proposes a systematic and straightforward approach to synthesize combinational digital logic with bimolecular reactions based on Karnaugh map. Boolean values' representation is the same as that in \cite{jiang2013digital}. However, when implementing combinational logic, this work focuses more on the intimate relationship between the function requirement and the design process. A direct link between the Karnaugh map and the reaction network is established, thus sophisticated circuit structure is saved. Such elaborate design discovers and exploits the natural programmability of biochemical systems in terms of digital logic, providing a much easier mapping procedure. Also, ODE simulation has ensured both accuracy and robustness. Table \ref{tz} lists all $2$-input logic gates synthesized by \cite{zhang2015karnaugh}. Its extended version shown in \cite{ge2017formal} offers more valid proofs from a mathematical perspective for the proposed approaches. Multi-input gates could be high efficiently synthesized with \cite{ge2017formal}. Figure \ref{fig:ge1} partially shows a 3-input gate implementation with CRNs. The corresponding simulation result is shown in Figure \ref{fig:ge2}.

\begin{table*}[htbp]
\centering
\setlength{\abovecaptionskip}{0pt}
\setlength{\belowcaptionskip}{4pt}
\caption{Simulation of all 2-input gates \cite{zhang2015karnaugh}.}
\centerline{
\begin{tabular}{cccc}
\raisebox{0\height}{\includegraphics[width=3.3cm]{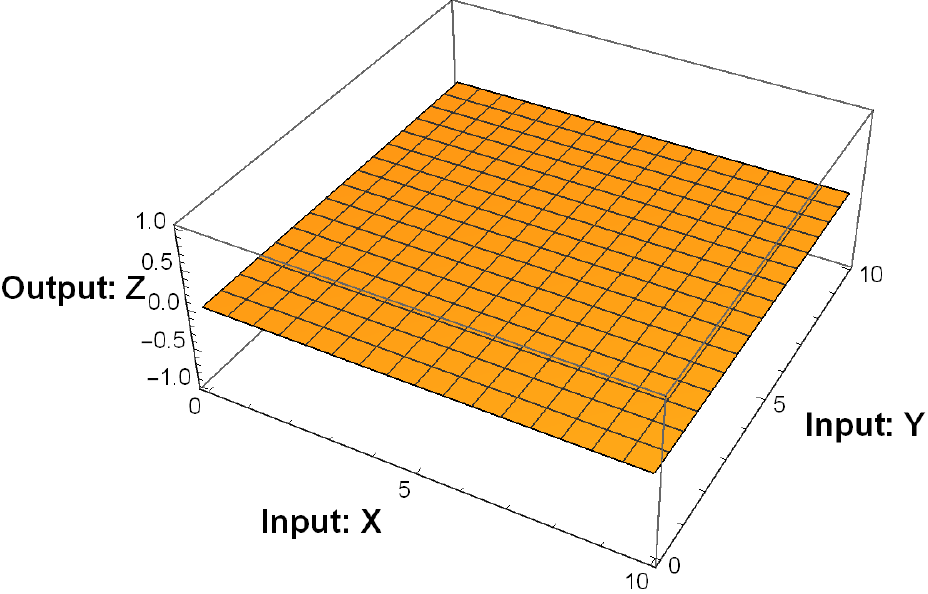}} & \raisebox{0\height}{\includegraphics[width=3.3cm]{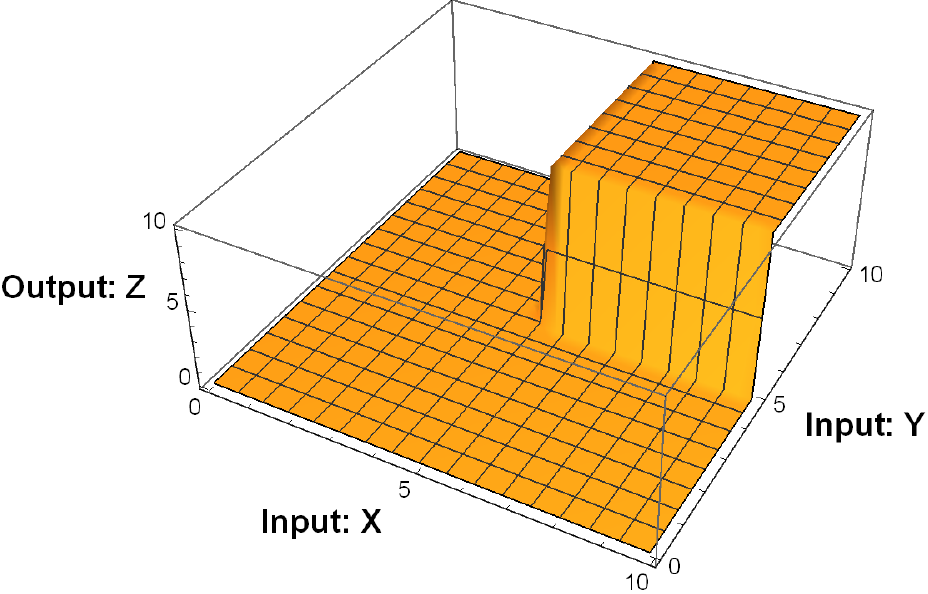}} & \raisebox{0\height}{\includegraphics[width=3.3cm]{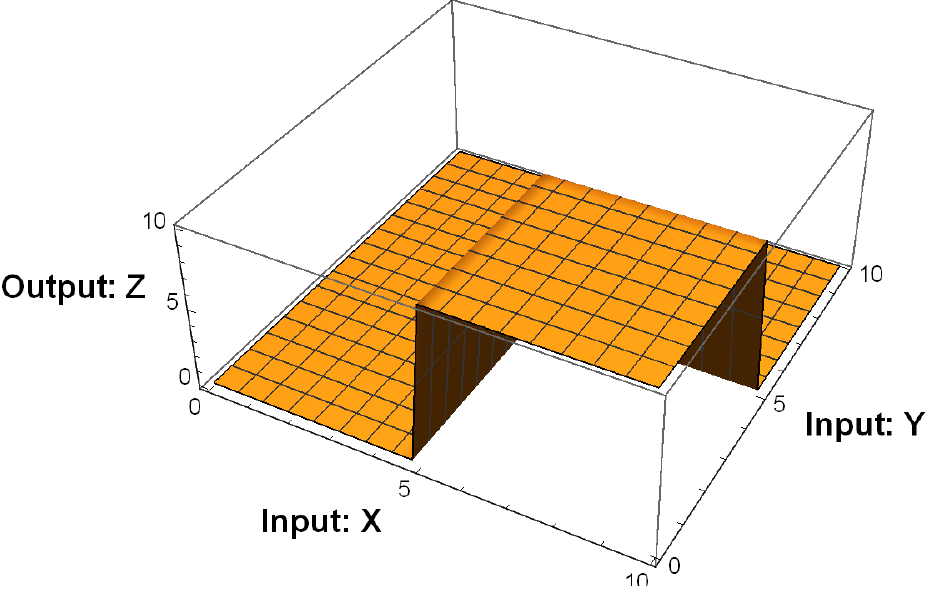}} & \raisebox{0\height}{\includegraphics[width=3.3cm]{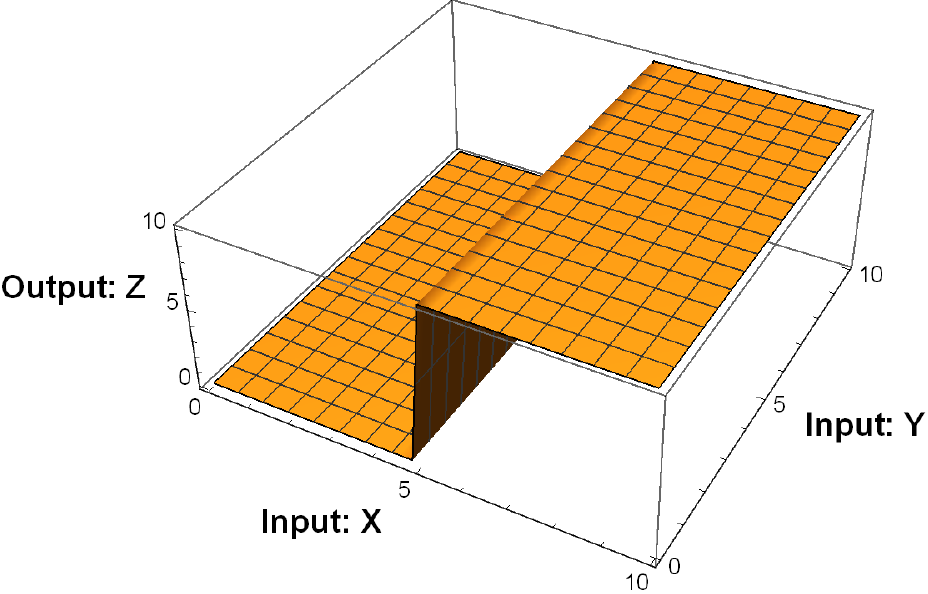}}\\ 
Side view of $[0000]$ & Side view of $[0001]$ & Side view of $[0010]$ & Side view of $[0011]$\\ 
 \raisebox{0\height}{\includegraphics[width=2.3cm]{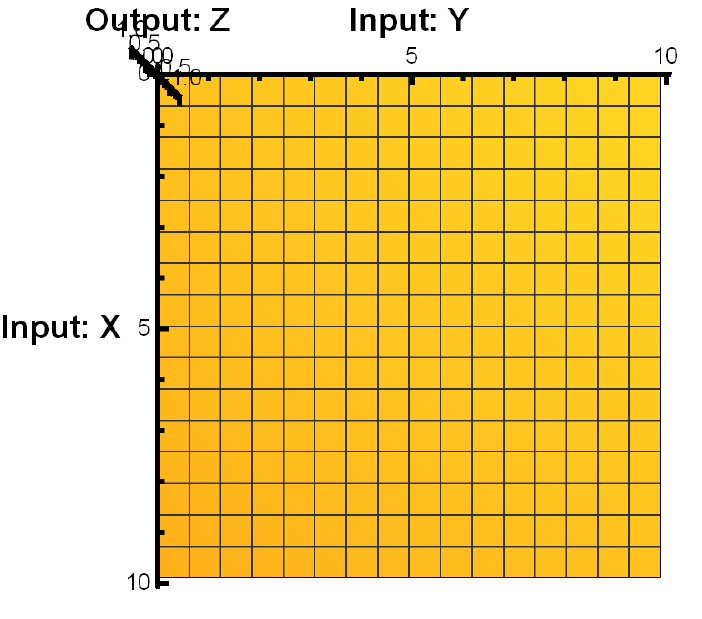}} & \raisebox{0\height}{\includegraphics[width=2.3cm]{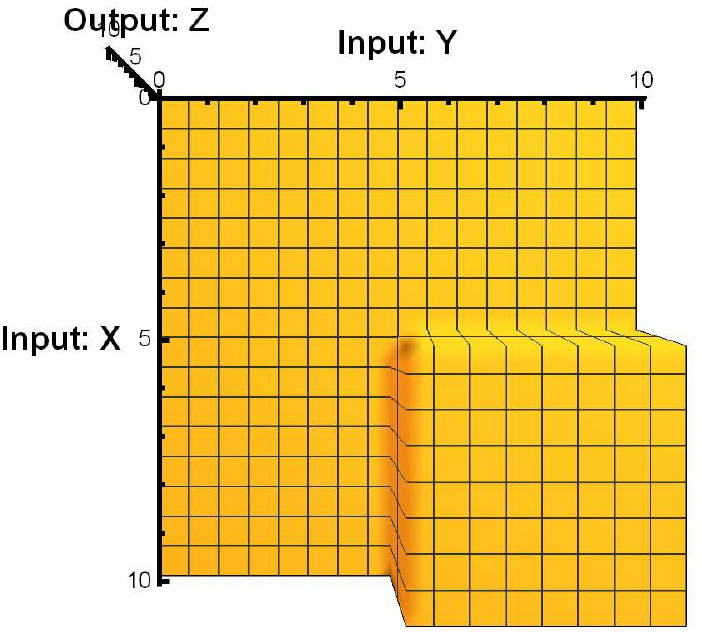}} & \raisebox{0\height}{\includegraphics[width=2.3cm]{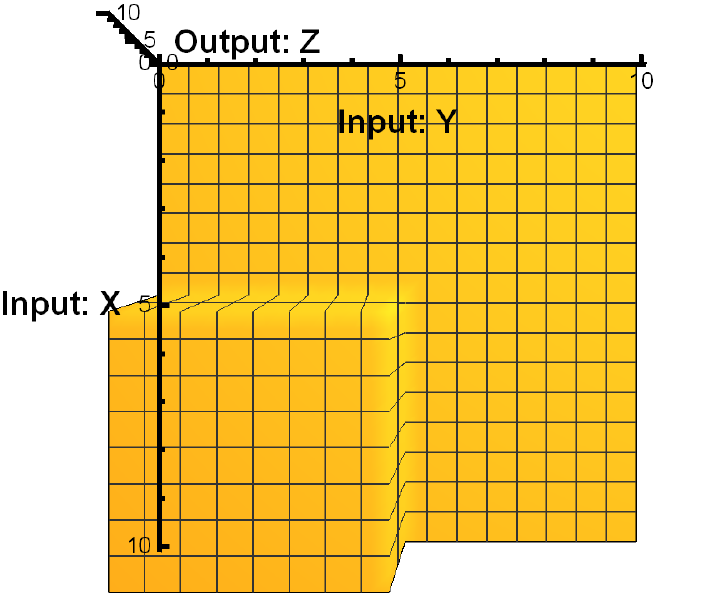}} & \raisebox{0\height}{\includegraphics[width=2.3cm]{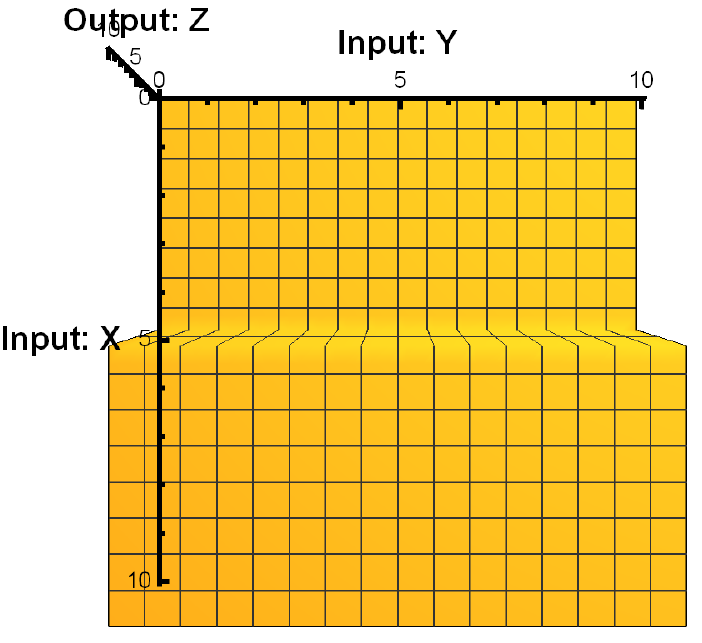}}\\
Top view of $[0000]$ & Top view of $[0001]$ & Top view of $[0010]$ & Top view of $[0011]$\\
\raisebox{0\height}{\includegraphics[width=3.3cm]{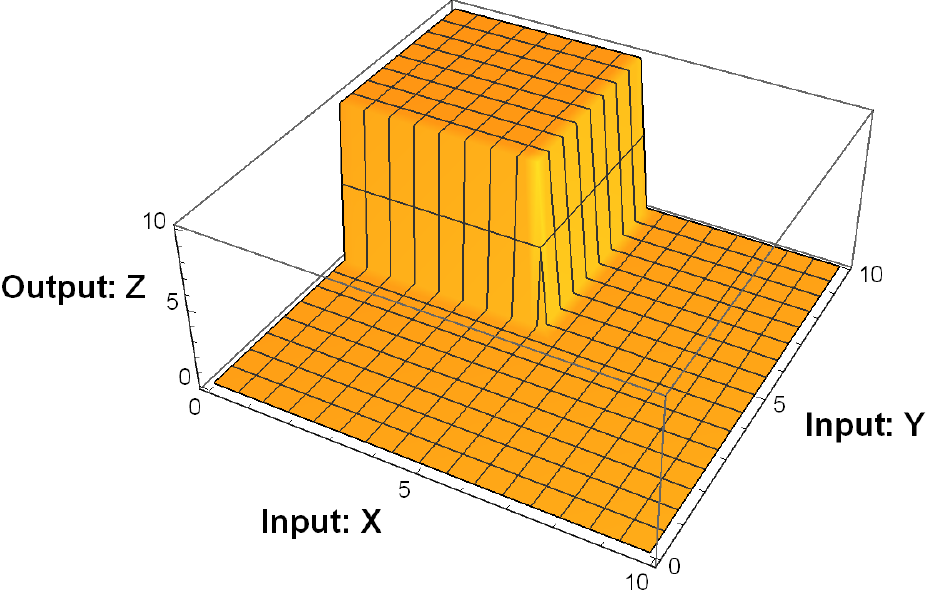}} & \raisebox{0\height}{\includegraphics[width=3.3cm]{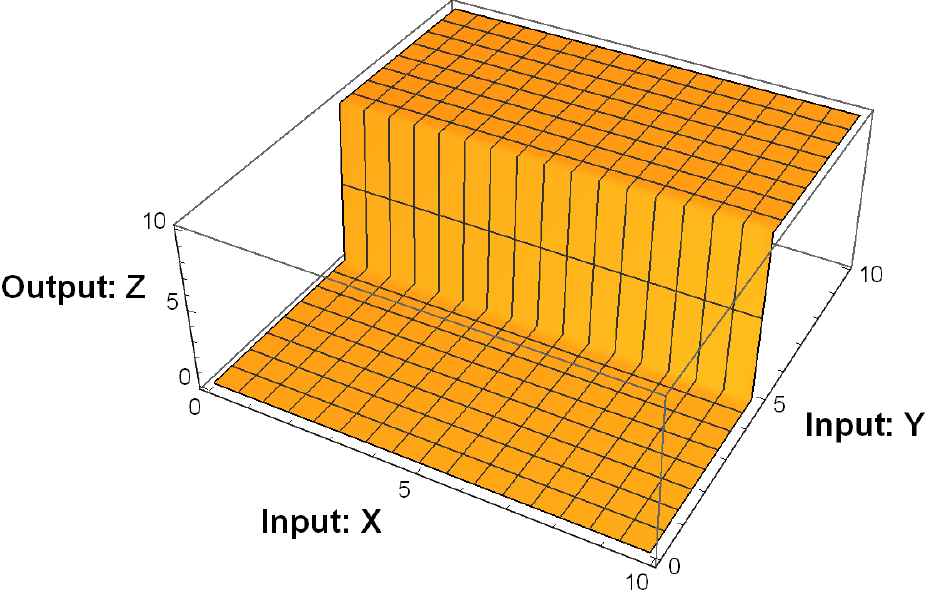}} & \raisebox{0\height}{\includegraphics[width=3.3cm]{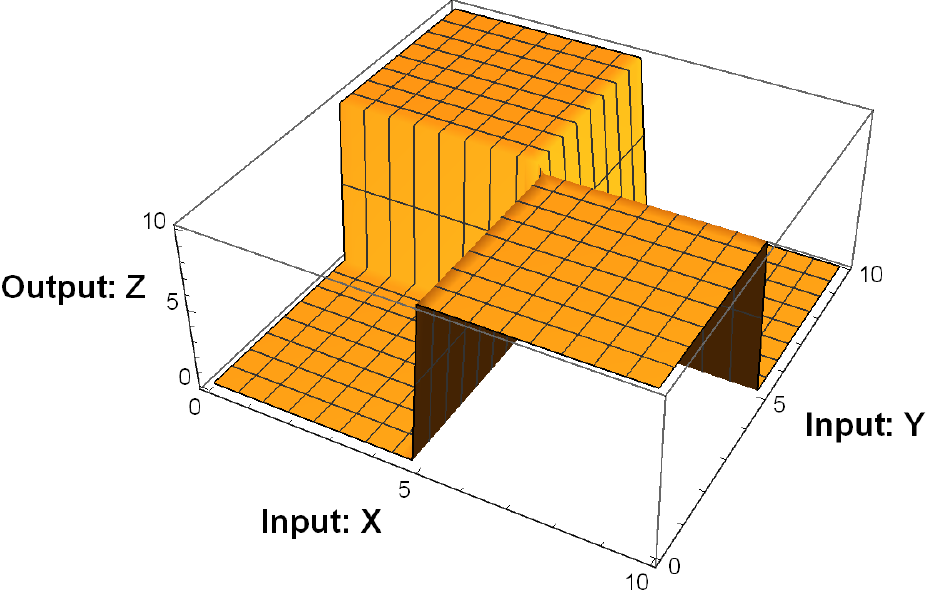}} & \raisebox{0\height}{\includegraphics[width=3.3cm]{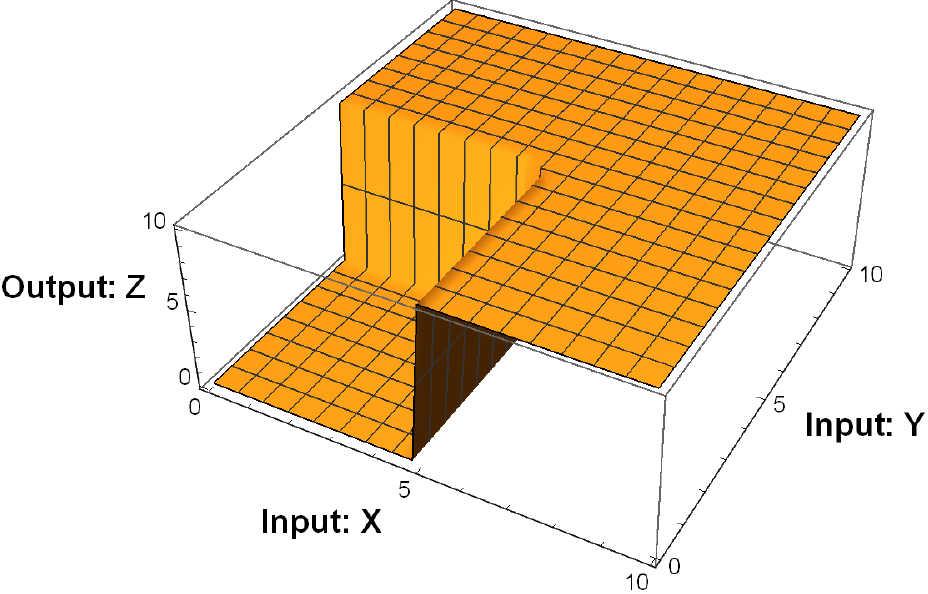}}\\ 
Side view of $[0100]$ & Side view of $[0101]$ & Side view of $[0110]$ & Side view of $[0111]$\\ 
 \raisebox{0\height}{\includegraphics[width=2.3cm]{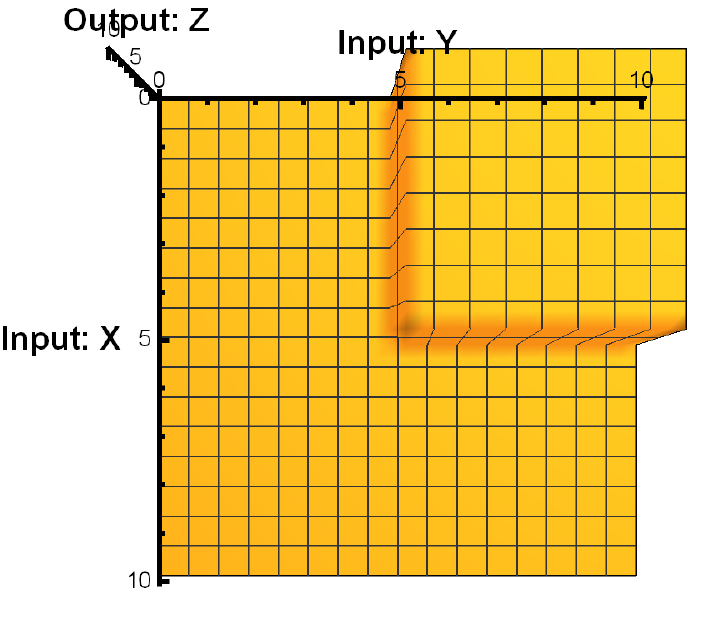}} & \raisebox{0\height}{\includegraphics[width=2.3cm]{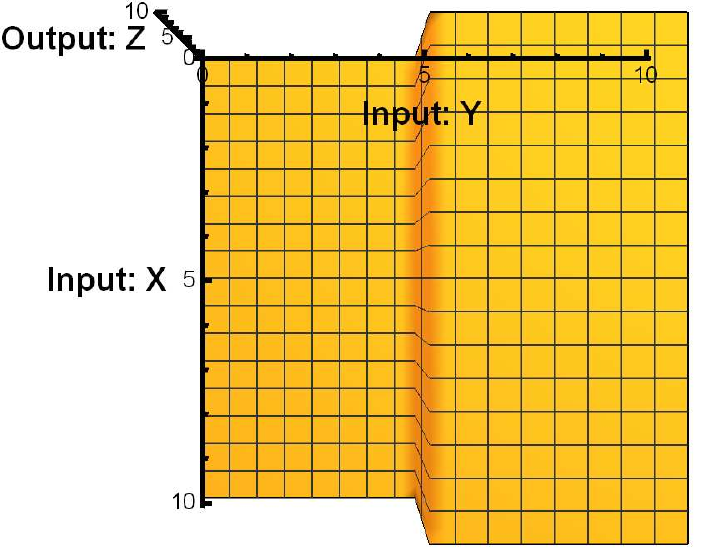}} & \raisebox{0\height}{\includegraphics[width=2.3cm]{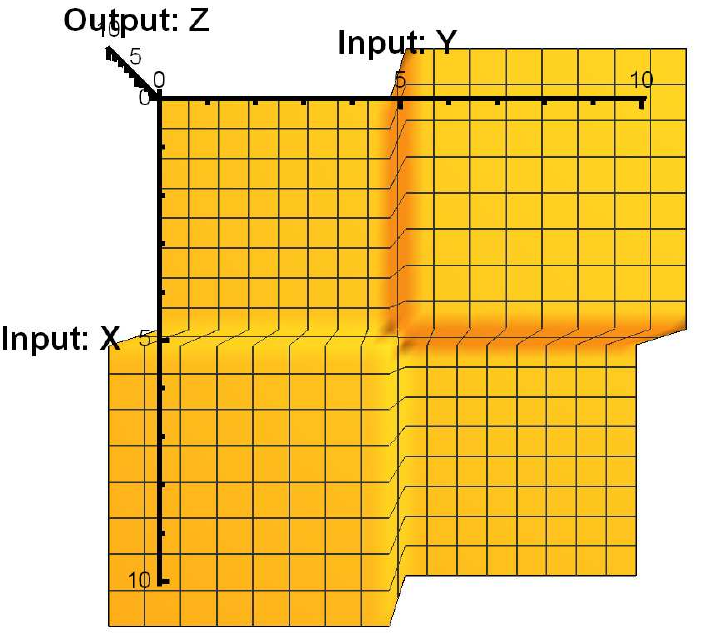}} & \raisebox{0\height}{\includegraphics[width=2.3cm]{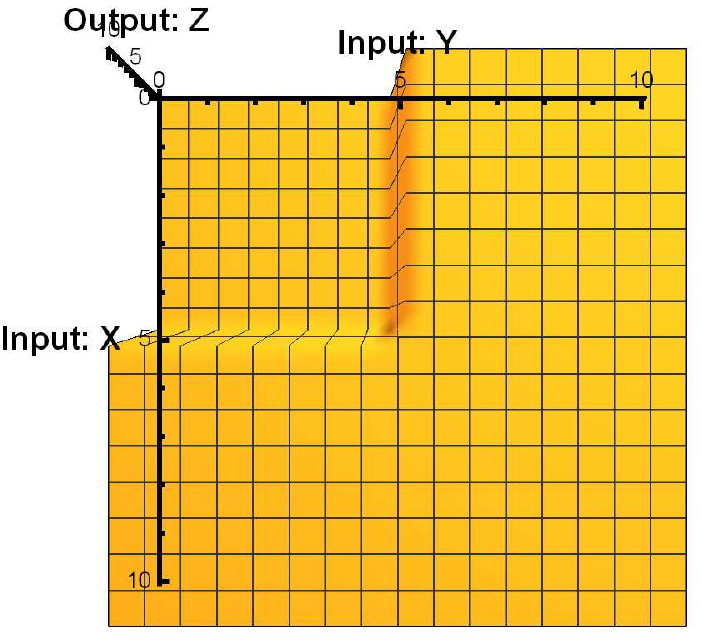}}\\
Top view of $[0100]$ & Top view of $[0101]$ & Top view of $[0110]$ & Top view of $[0111]$\\
\raisebox{0\height}{\includegraphics[width=3.3cm]{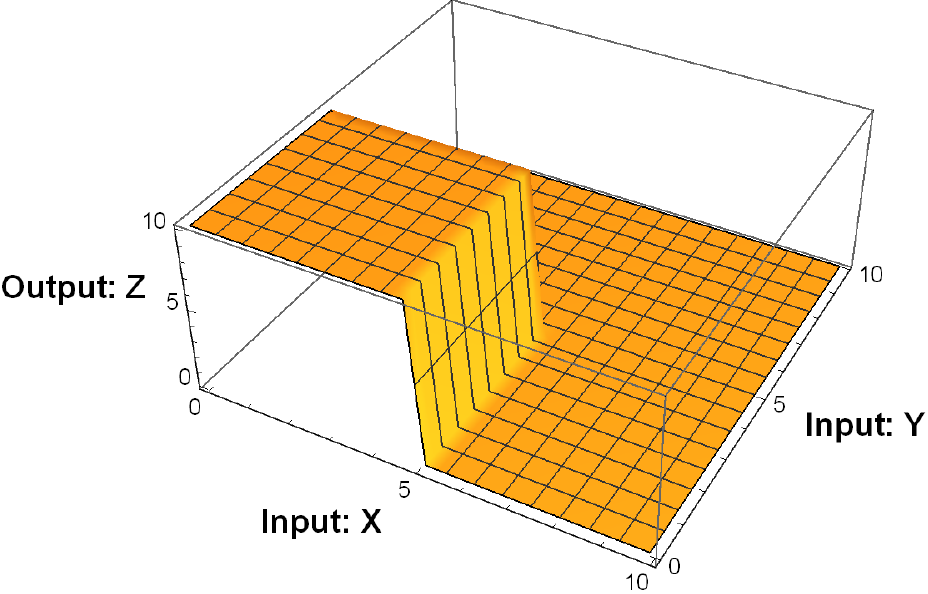}} & \raisebox{0\height}{\includegraphics[width=3.3cm]{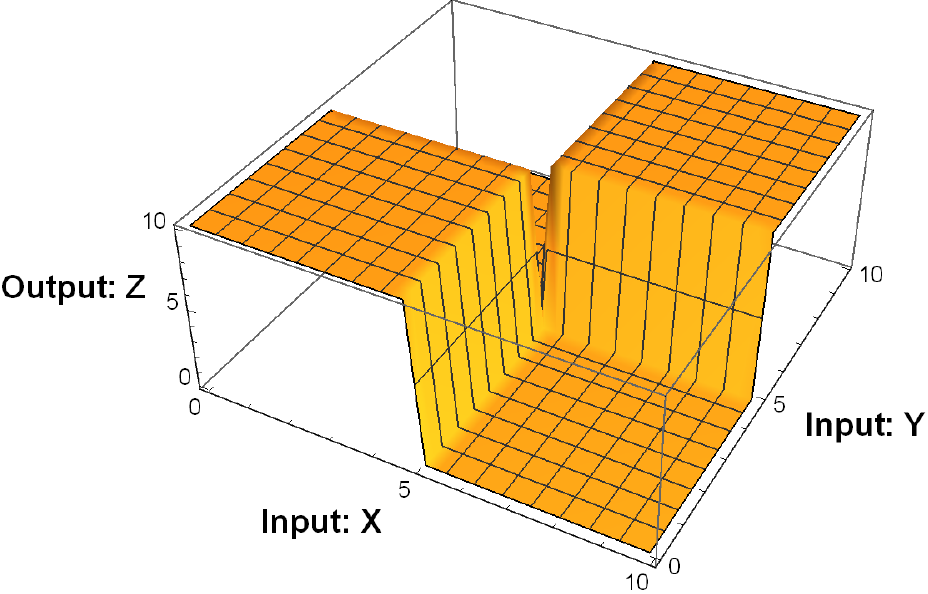}} & \raisebox{0\height}{\includegraphics[width=3.3cm]{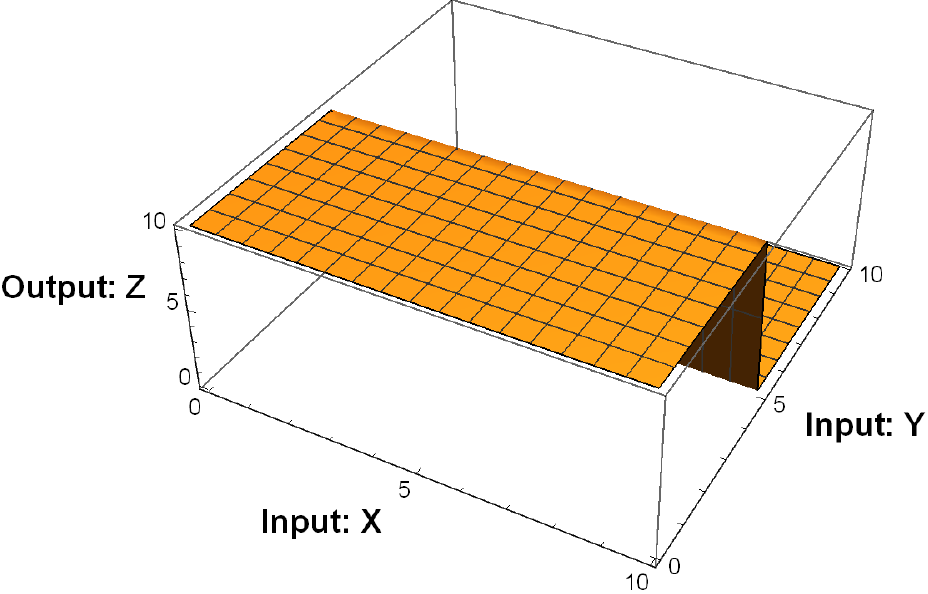}} & \raisebox{0\height}{\includegraphics[width=3.3cm]{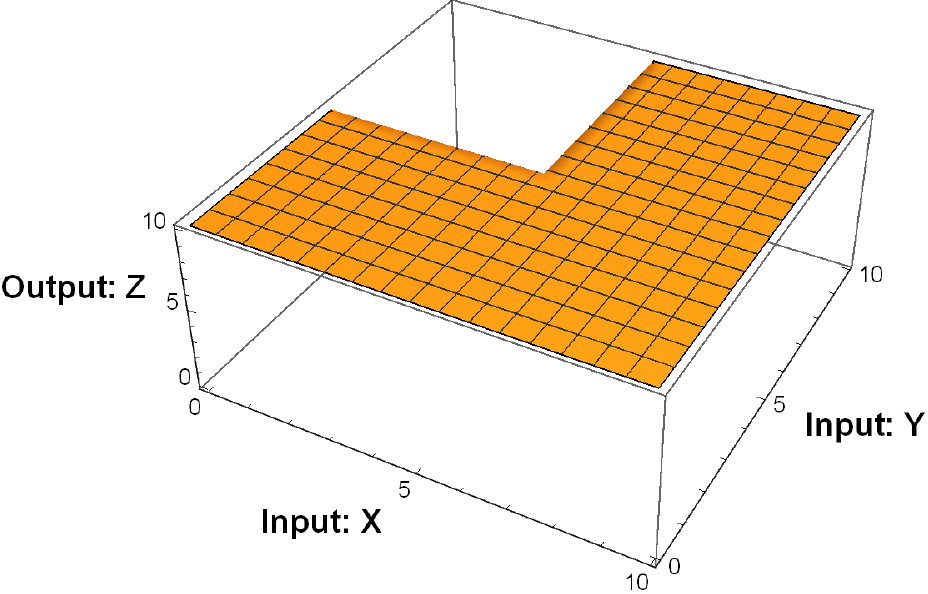}}\\ 
Side view of $[1000]$ & Side view of $[1001]$ & Side view of $[1010]$ & Side view of $[1011]$\\ 
 \raisebox{0\height}{\includegraphics[width=2.3cm]{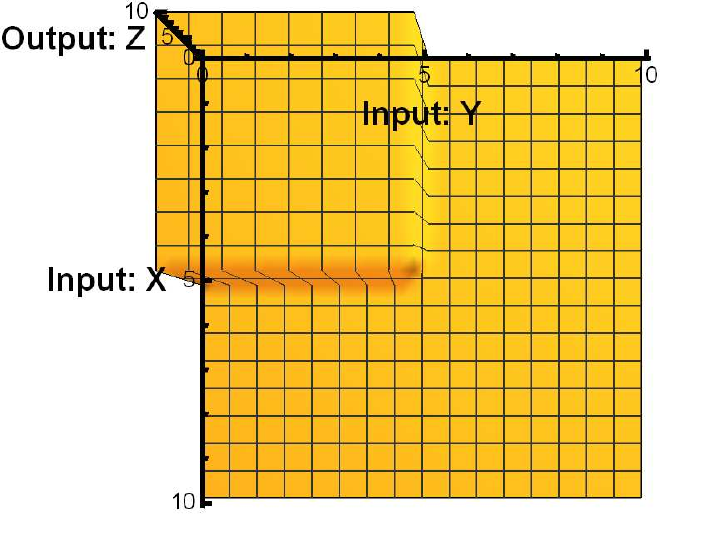}} & \raisebox{0\height}{\includegraphics[width=2.3cm]{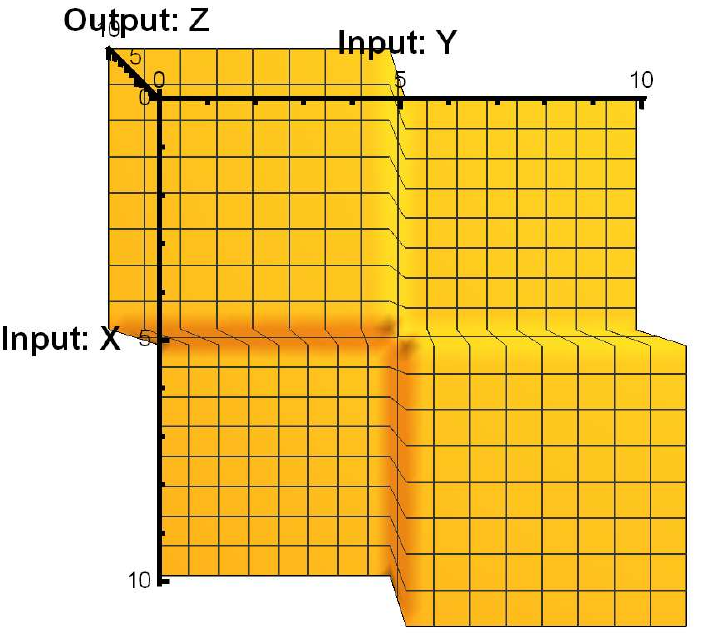}} & \raisebox{0\height}{\includegraphics[width=2.3cm]{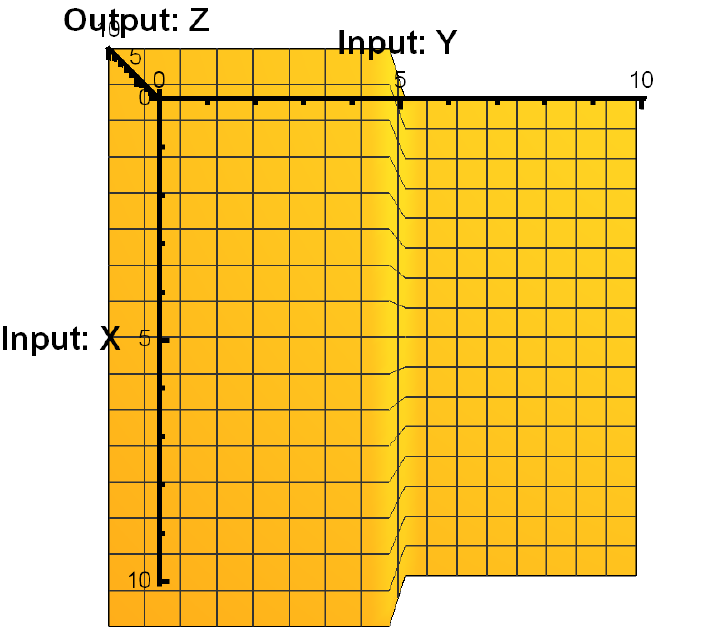}} & \raisebox{0\height}{\includegraphics[width=2.3cm]{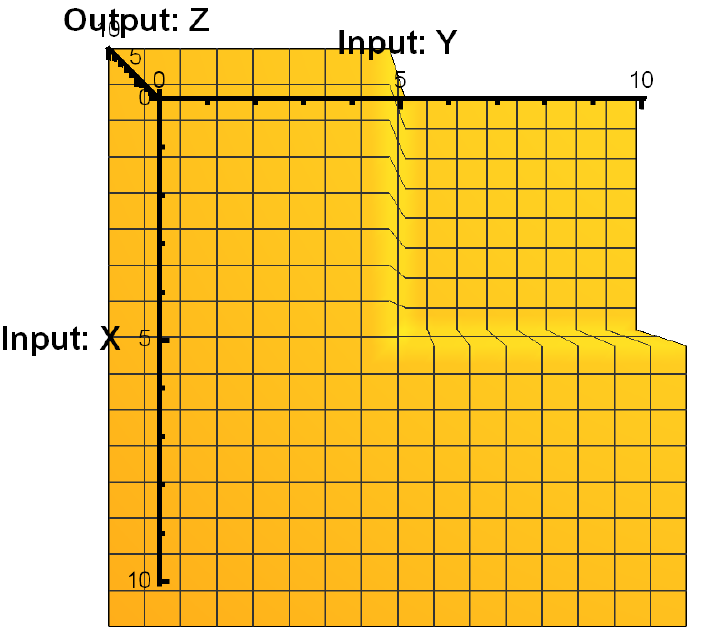}}\\
Top view of $[1000]$ & Top view of $[1001]$ & Top view of $[1010]$ & Top view of $[1011]$\\
\raisebox{0\height}{\includegraphics[width=3.3cm]{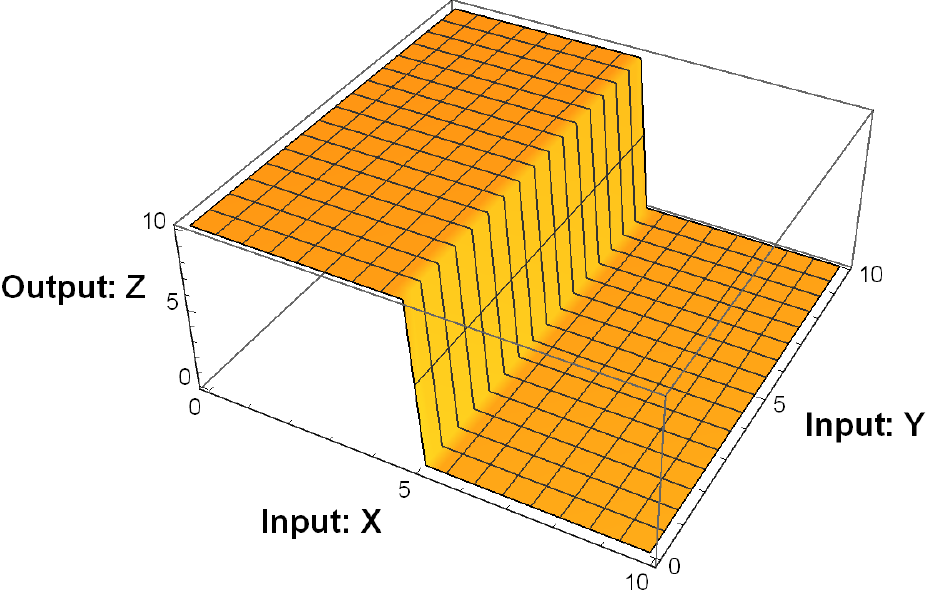}} & \raisebox{0\height}{\includegraphics[width=3.3cm]{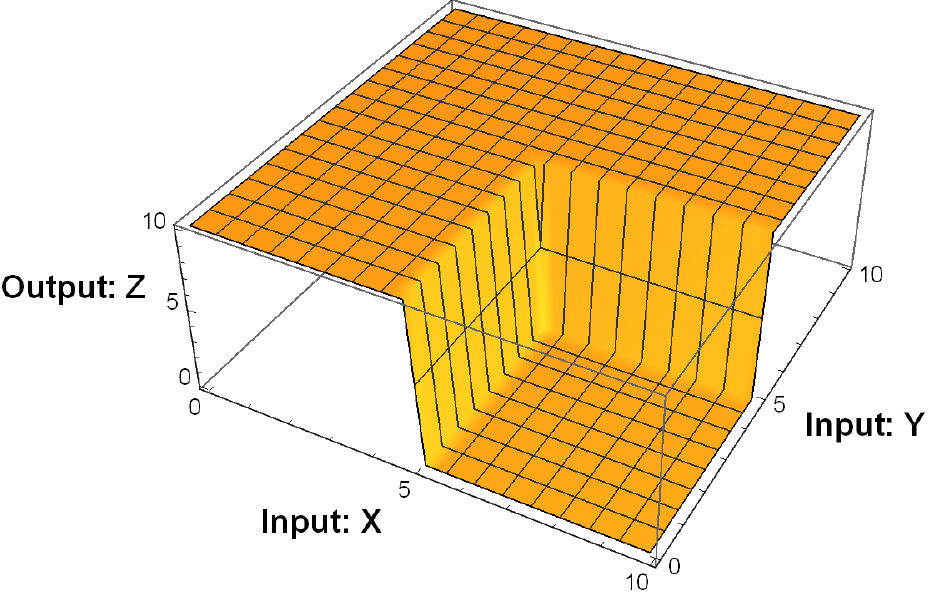}} & \raisebox{0\height}{\includegraphics[width=3.3cm]{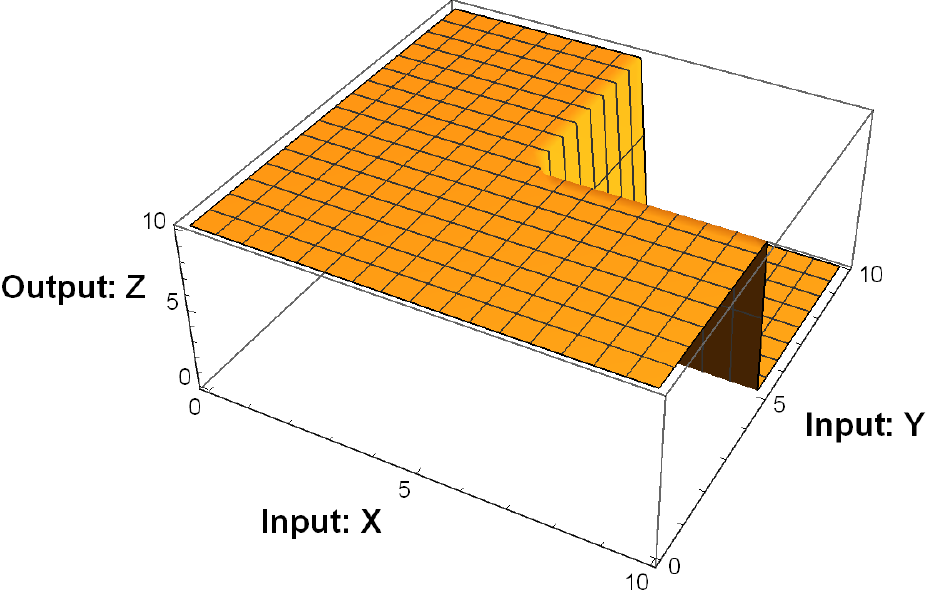}} & \raisebox{0\height}{\includegraphics[width=3.3cm]{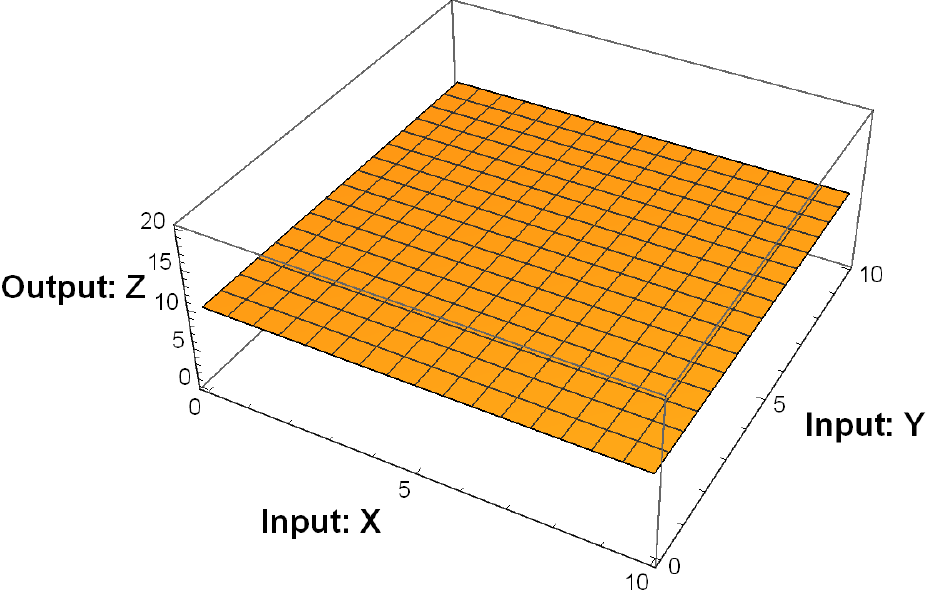}}\\ 
Side view of $[1100]$ & Side view of $[1101]$ & Side view of $[1110]$ & Side view of $[1111]$\\ 
 \raisebox{0\height}{\includegraphics[width=2.3cm]{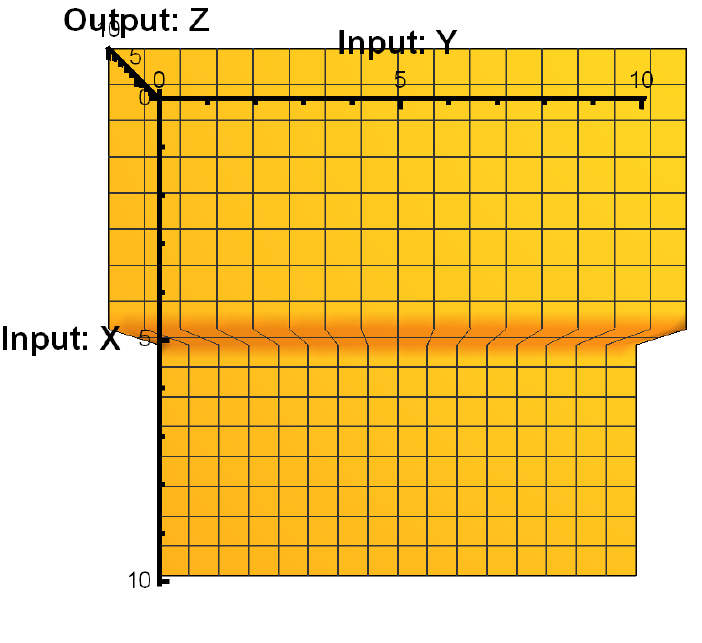}} & \raisebox{0\height}{\includegraphics[width=2.3cm]{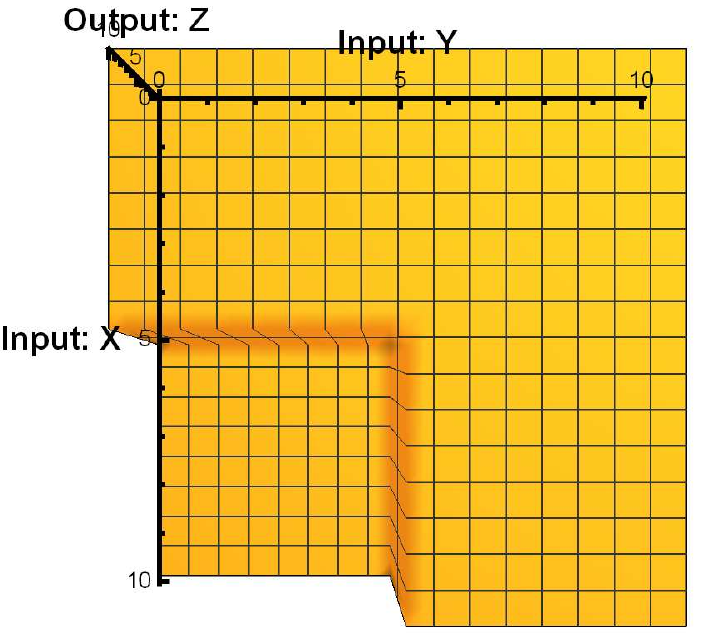}} & \raisebox{0\height}{\includegraphics[width=2.3cm]{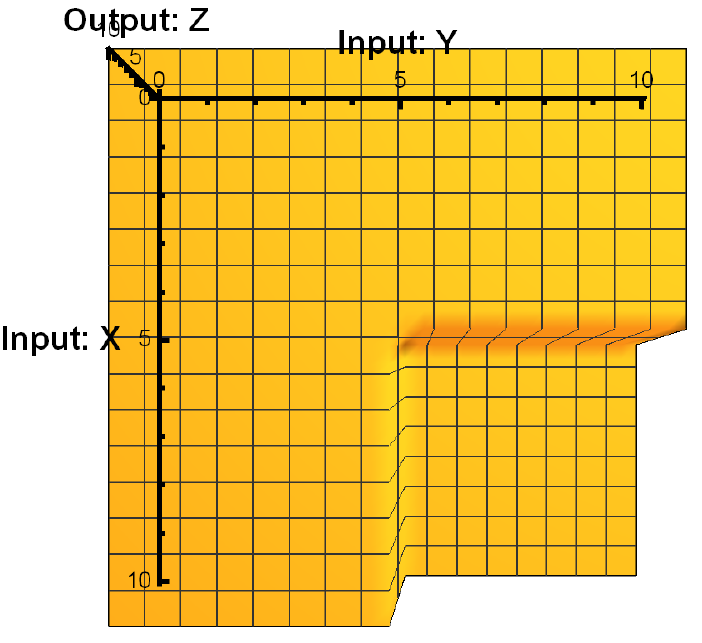}} & \raisebox{0\height}{\includegraphics[width=2.3cm]{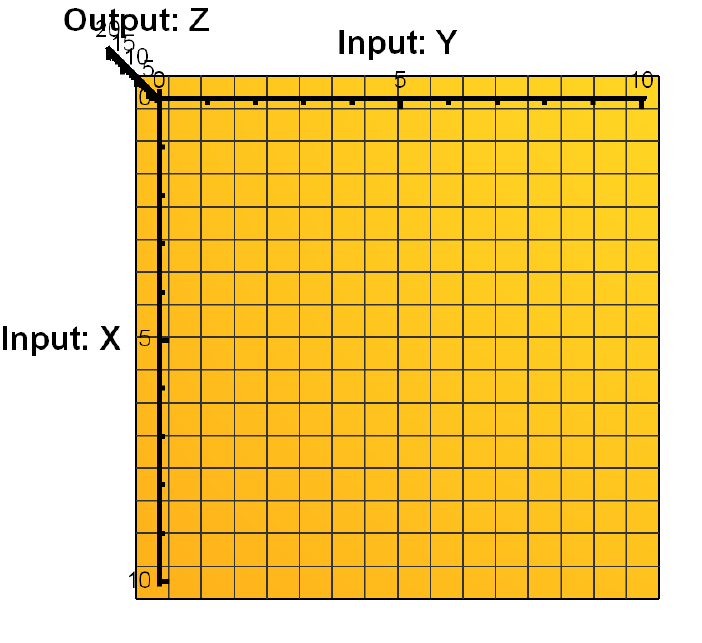}}\\
 Top view of $[1100]$ & Top view of $[1101]$ & Top view of $[1110]$ & Top view of $[1111]$\\
\end{tabular}
}
\label{tz}
\end{table*}

In addition, \cite{zadegan2015construction} proposes a design approach for the operation of all the Boolean logic gates, including AND, NAND, OR, NOR, XOR and XNOR operations. Beyond that, a fuzzy logic gate system, as a proof-of-concept, is also successfully implemented in a DNA origami box. These basic Boolean DNA logic gates are all designed from a universal structure, namely two types of DNA complexes, and can be readily constructed into larger scale. In this study, the input signals are typically DNA strands, while the output are denoted by fluorophores resonance energy transfer (FRET) signals. The proposed DNA-based logic gates can be combined to perform simple yes/no operations in a sensor, and would enable the calculations in vivo as well as cancer diagnostics at the molecular level.

 \begin{figure}[H]
\begin{minipage}[t]{1\linewidth}
\centering
\includegraphics[width =.7\linewidth]{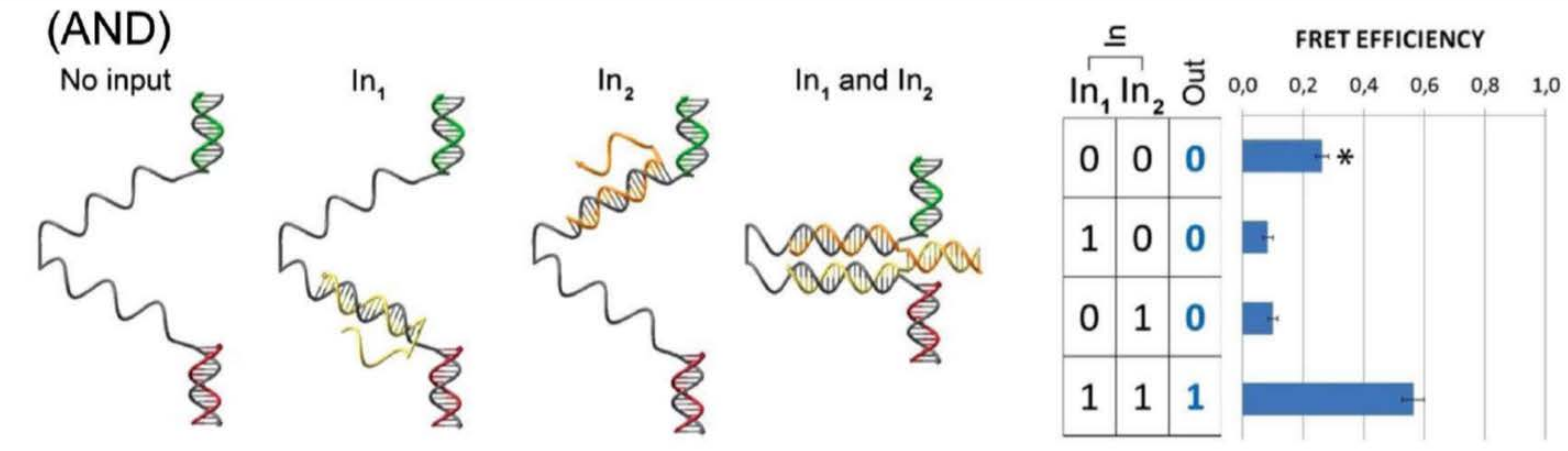}
\caption{Design principles of the AND gate based on DNA strands in  \cite{zadegan2015construction}.}
\end{minipage}%
\end{figure}

\section{Future research}\label{sec:future}
\quad As a non-von Neumann architecture machine, DNA computer tries to rethink the notion of computation. Operating in a parallel manner, it starts all the bio-operations from encoding information over the alphabet $\{A,T,G,C\}$. Just like the silicon-based computer performs complex calculations on a basic suite of arithmetic and logical operations, DNA has cutting, copying and many others. ``Born'' to solve a seven-point Hamiltonian path problem, DNA computing shows overwhelming advantages in solving hard problems, especially those problems that conventional computers cannot address. Seemingly as the smallest computer, about a kilogram of its computational substrate DNA could meet the world's storage requirements as long as the information could be packaged densely enough. Due to certain limitations, there is a current consensus that DNA computing will not be the rival for silicon-based one. This does not indicate that DNA computing is destroyed in the water. Most likely, the rich potential of DNA computing lies in \textit{vivo} computing as well as in being combined with existing machinery to form DNA/silicon hybrid architectures. A future is predicted, in which our bodies are patrolled by a lot of tiny DNA computers, whose tasks are to monitor our well-being and release the right drugs to repair damaged or unhealthy tissue. Automatization is also potential, freeing most bio-operations from manual handling of tubes. Expected to be good at doing DNA-related things, DNA computing might be used in information storage, encryption, medical diagnosis, drug delivery and so on and so forth. To develop DNA computing in its full motion, there is still a long way off, full of both hopefulness and challenges.

As the core components of digital circuits, combinational logic gates could be synthesized with DNA strands to perform basic programmable logical operations. They as well could be cascaded to form more complex logic circuits---including sequential logic---and even larger systems to realize complex tasks and bio-sensing. Employing combinational logic gates is essentially an approach belonging to digital field, while analog signals are more common in nature. Thus, a new trend calling for more efficient DNA computation is to mix analog and digital computation together. To this end, both the molecular analog-to-digital converters and digital-to-analog converters are heated in the study of mixing these two forms of computation in the long run. In addition, there is a bold assumption or application beyond basic combinational logics, that is to implement fuzzy logic, probabilistic logic systems and even a Turing machine. They operate with analog logic where inputs and outputs have values between 0 and 1 or partial truth statements \cite{zadegan2015construction}. To elaborate a little further, unlike the deterministic interpretation of a truth table we are familiar with, a value is interpreted as the degree of truth to what extent a proposition is true, or the probability that the proposition is true. The importance of fuzzy logic arises from its role in both academia and industry, covering the field of medical diagnosis, medicine delivery \cite{lu2013dna,li2011self}, artificial neural networks \cite{qian2011neural,schneider1998artificial,noordewier1991training}, RNA secondary structure prediction \cite{zuber2017sensitivity}, image understanding \cite{zhang2017tissue}, processing by artificial intelligence \cite{brady1985artificial}, and fuzzy electronics \cite{ray2011similarity,jeng2006fuzzy,zadegan2015construction}. Therefore, in order to enhance our understanding of natural systems and additionally provide a new tool to perform soft computing, it is important for us to understand fuzzy logic and apply it to artificial synthetic biology, especially in combination with DNA origami.



\Acknowledgements{This work is partially supported by NSFC under grant $61501116$, Jiangsu Provincial NSF under grant BK$20140636$, Huawei HIRP Flagship under grant YB$201504$, International Science \& Technology Cooperation Program of China under grant $2014$DFA$11640$, ICRI for MNC, the Fundamental Research Funds for the Central Universities, the Project Sponsored by the SRF for the Returned Overseas Chinese Scholars of State Education Ministry, and Student Research Training Program of Southeast University.}

\end{document}